\documentclass[%
 reprint,
 amsmath,amssymb,
 aps,prl,
]{revtex4-2}
\usepackage{paralist}
\usepackage{graphicx}
\usepackage{tabularx}
\usepackage{dcolumn}
\usepackage{bm}
\usepackage{color}
\usepackage{etoolbox}
\usepackage{tikz}
\usepackage{ulem}

\newrobustcmd*{\mysquare}[1]{\tikz{\filldraw[draw=#1,fill=#1] (0,0)
rectangle (0.2cm,0.2cm);}}

\newrobustcmd*{\mycircle}[1]{\tikz{\filldraw[draw=#1,fill=#1] (0,0) circle [radius=0.1cm];}}

\newrobustcmd*{\mytriangle}[1]{\tikz{\filldraw[draw=#1,fill=#1] (0,0) --
(0.2cm,0) -- (0.1cm,0.2cm);}}

\newrobustcmd*{\myinvtriangle}[1]{\tikz{\filldraw[draw=#1,fill=#1] (0.1cm,0) --
(0.2cm,0.2cm) -- (0.cm,0.2cm);}}

\usepackage{enumitem}

\newcommand{\pder}[2]{\partial_{#1}{#2}}
\newcommand{\pdder}[3]{\partial_{#1}\partial_{#2}{#3}}
\newcommand{\figref}[1]{Fig.~\ref{#1}}

\def\YD#1{\textcolor{black}{#1}}

\begin{document}



\preprint{APS/123-QED}

\title{A first coherent structure in elasto-inertial turbulence}

\author{Y. Dubief$^1$}\email{ydubief@uvm.edu}
 
\author{J. Page$^2$}%
\author{R. R. Kerswell$^2$}
\author{V. E. Terrapon$^3$}
\author{V. Steinberg$^4$}
\affiliation{ $^1$ Department of Mechanical Engineering, University of Vermont, Burlington, VT, USA\\
$^2$ Department of Applied Mathematics  and Theoretical Physics,  University of Cambridge, Cambridge, UK\\
$3$ Aerospace and Mechanical Engineering Department,
 University of Liege,
 Belgium\\
$^4$ Department of Physics of Complex Systems, Weizmann Institute of Science, Rehovot, Israel
}%

\date{\today}

\begin{abstract}
Two dimensional channel flow simulations of FENE-P fluid in the elasto-inertial turbulence regime reveal distinct regimes ranging from chaos to 
a steady travelling wave which takes the form of an arrowhead structure.  This coherent structure provides new insights into the polymer/flow interactions driving EIT. A set of controlled numerical experiments and the study of transfer between elastic and turbulent kinetic energies highlight the role of small- and large-scale dynamics in the self-sustaining cycle of chaos in EIT flows.   
\end{abstract}

\maketitle

\section{Introduction}
Elasto-inertial turbulence (EIT) \cite{Samanta2013el,Dubief2013hh} is a chaotic state occurring in 
weakly inertial to strongly inertial channel and pipe flows with polymer additives. Given an appropriate initial perturbation, local fluctuations of velocity gradient stretch polymers, which exert a local stress feedback on the flow, thereby sustaining a level and an organization of velocity gradient fluctuations. The exact mechanism of  interaction betwen flow instabilities and polymer instabilities remains poorly understood. EIT belongs to the category ``active scalar turbulence'' where a molecule (\textit{e.g.} polymers), an organism (\textit{e.g.} bacteria, microswimmers) or a field (\textit{e.g.} magnetic field) is two-way coupled with the flow and this coupling has a direct influence on, or may drive, turbulence. The present study is concerned with a sub-category of active turbulence, that of additive-driven chaos. The dynamics is a self-sustaining cycle  where the chaotic dynamics of the scalar induces flow perturbations that in turns sustain the scalar's chaotic dynamics. Such cycles are observed in active turbulence induced by bacteria or microswimmers \cite{wensink2012meso} or nematic fluids \cite{alert2020universal}, at very low Reynolds number. Also in inertialess flows, and relevant to the present study, polymer additives create elastic turbulence \cite{groisman2000elastic,groisman2001efficient} in flows with curved streamlines. 

EIT, elastic turbulence, active nematic turbulence or active turbulence induced by bacteria promote mixing in flows either dominated by diffusion or where Newtonian instabilities cannot survive in the absence of the active scalar. This chaos has practical applications, such as heat transfer enhancement \cite{traore2015efficient} or promoting emulsification \cite{poole2012emulsification} for elastic turbulence. From a theoretical perspective, these flows are fundamentally different from classical turbulence, specifically with respect to energy transfers between large and small scales. In the case of EIT, a numerical experiment consisting of increasing the molecular diffusivity of the polymer model demonstrated that the small-scale dynamics of polymers is critical to sustaining chaos \cite{sid2018two}. Our ability to harness and control the mixing power of such an active scalar ultimately requires the understanding of how polymer and flow scales interact, and of how the flow transfers energy to the active scalar and \textit{vice versa}. These energy transfers, in turbulent systems where they are identified, are directly linked to coherent structures. For instance, the self-sustaining dynamics of wall-bounded turbulence relies on quasi-streamwise vortices which interact with streaks, elongated regions of high- and low-momentum fluid, to create vertical energy transfer to and from the very near wall regions \cite{robinson1991coherent,jimenez1999autonomous}. 

The notion of coherent structures is ubiquitous in the theory of turbulence, even though its definition remains empirical. A coherent structure is typically defined as a region of the flow whose dynamics (a) has a significant energetic impact and (b) whose dynamics remain correlated for sufficiently large time scales, at least larger that the smallest time scale of turbulence, defined by Kolmogorov's  theory \cite{kolmogorov1941local} in classical turbulence. So far, the study of EIT has revealed structures in the form of thin, elongated sheets where polymers are much more stretched than anywhere else in the flow \cite{Samanta2013el}. Attached to these sheets are trains of spanwise, cylindrical regions of positive and negative $Q:=\YD{-}\tfrac{1}{2} \partial_j u_i \partial_i u_j$, the second invariant of the  velocity gradient tensor ($\pder{j}{u_i}$ is the $j$-derivative of the velocity component $i$) \cite{dubief2010polymer,Terrapon2014wu}. $Q$ is the basis for a common vortex identification method in classical turbulence \cite{dubief2000coherent}, as it is the difference between the local norm of the rotation rate and the strain rate. It is also related to local minima of pressure through the Laplacian of pressure $2Q=\pdder{i}{i}{p}$. However, in EIT flows, the regions of positive $Q$, where the local rotation rate is larger than the local strain rate, are only strong enough to produce oscillations in the local streamlines rather than vortices \cite{dubief2010polymer}.

Implied in the above description of the structure of $Q$ is the hypothesis that the existence of thin polymer sheets excites elastic flow instabilities, akin to but distinct from the Kelvin-Helmholtz instability in shear flows. One equation central to this hypothesis is the elliptic equation for pressure
\begin{equation}
    \pdder{i}{i}{p}=2Q+\frac{1-\beta}{Re}\pdder{i}{j}{T_{ij}} \,,\label{eq:ddp}
\end{equation}
where $p$ is the pressure, and  $T_{ij}$ and $(1-\beta)/Re$ are the polymer stress tensor and viscosity parameters related to the polymer solution, both of which will be defined later. This equation is the result of applying the divergence operator to the momentum transport equation. Equation~\eqref{eq:ddp} is also believed to be a key equation of elastic turbulence \cite{burghelea2007elastic} as it connects polymer dynamics, pressure and the nonlinear inertial effect ($Q$ is the divergence of the advection term in the Navier-Stokes equation). Note that this last term is small compared to the other terms in the regimes of low Reynolds number considered for this study, even at $Re={\cal{O}}(1000)$.

Due to the elliptic nature of Eq.~\eqref{eq:ddp}, it can be anticipated that small scale perturbations in polymer stress, amplified by the second spatial derivative, have an instantaneous, local and global effect on the pressure. Local variations of pressure would in turn translate into velocity perturbations and creation of local strain, which drives polymer stretching.
Recently, it has also been argued that structures connected to Newtonian Tollmien-Schlicting instability waves may play a role in EIT \cite{ashwin2019critical}.
There is numerical evidence that weakly chaotic states, which arise 
in a sequence of bifurcations from the Newtonian travelling waves, can be continued into regions of the parameter space where EIT has also been observed. 
However, this work has been restricted to very dilute solutions with
weak polymer activity and it is currently unclear whether the polymer sheets are the cause or the effect of the relevant instabilities in EIT.

The present study reports the discovery of a first coherent structure in EIT which can dominate the dynamics of the flow. Under certain conditions, the flow becomes steady and symmetrical about the mid-plane of the channel with the coherent structure being the only, but significant, departure from a laminar flow.  Based on the structure of the associated  polymer stress, this structure looks like an arrowhead which points in the direction of the flow and propagates at a constant speed downstream. Under certain conditions, the `arrowhead' may become a robust attractor with complete elimination of chaos. At the macroscopic level, the presence of the resulting travelling wave shows in a steady drag increase. 

The very existence of the structure is particularly exciting for a turbulence that does not  follow the classical properties of Newtonian turbulence. Power spectra show steeper decay of energy \cite{Dubief2013hh} than the $-5/3$ decay of the classical cascade of energy \cite{kolmogorov1941local}. The numerical experiments of Sid \textit{et al.} \cite{sid2018two} showed that EIT can only be sustained if the small scale dynamics of polymer stress is accurately captured in simulation, strongly suggesting an inverse energy cascade from polymer to flow perturbations. Due to its chaotic nature, analyses of EIT have been necessarily statistical in the absence of any simpler manifestation of the phenomenon. 
In statistics, it is often difficult to conclusively isolate the fine details of complex dynamics. The arrowhead -- the first coherent structure to be isolated from EIT --  provides a far simpler investigative framework thanks to its steadiness in an appropriately travelling frame. 
The goals of the present study are to introduce the existence of the  arrowhead structure and to identify and compare the energy transfer between polymers and flow for it and for the otherwise chaotic structures of EIT.

\section{Methods} 

\begin{figure*}[ht]
\centerline{\includegraphics[width=\textwidth]{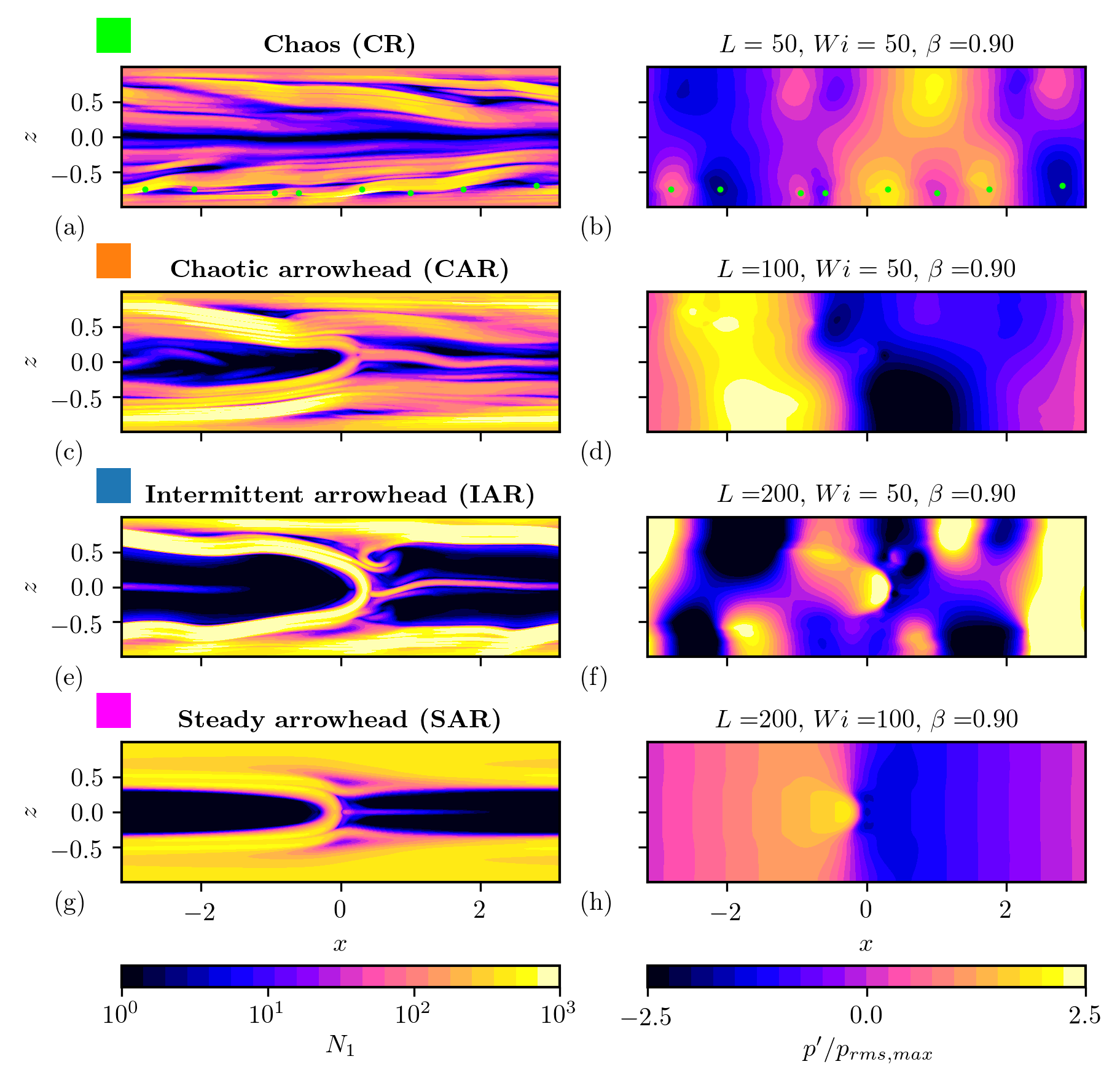}}
\caption{\label{fig:1}  Characteristic snapshots of the four distinct regimes: CR, chaotic regime (a,b), CAR chaotic arrowhead regime (c,d), IAR, intermittent arrowhead regime (e,f) and SAR, steady arrowhead regime (g,h), identified in the $L_x=2\pi$-computational domain. The left column shows contours of first normal stress difference on a log-scale colormap, the right column provides the corresponding pressure contours on a linear scale. In Figs. (a,b), the green dots identify the location of low and high pressure regions in the lower-half of the channel.}
\end{figure*}

The simulations discussed here solve the FENE-P (finitely extensible nonlinear elastic-Peterlin) viscoelastic model in 2D using the  algorithm \cite{dubief2005nai} used to discover EIT \cite{dubief2010polymer,dubief2012analysis,Samanta2013el,Dubief2013hh,Terrapon2014wu}. The computational domain is a channel with periodic boundary conditions over a length of $2\pi n_x h$ ($n_x=1$ for most simulations discussed here) in the streamwise direction $x=x_1$ and walls at $z=x_2=\pm h=\pm 1$. \YD{Note that graphs of wall-normal profiles of statistical quantities are shown as function of the distance from the wall,}
\begin{equation}
    \xi = h-z\,.
\end{equation}
The flow is divergence free, $\pder{i}{u_i}=0$, where $u_i$ is the velocity vector. The momentum transport equation is
\begin{equation}
    D_t{u_i}=-\pder{i}{p}+\frac{\beta}{Re}\pdder{j}{j}{u_i}+\frac{1-\beta}{Re}\pder{j}{T_{ij}} + f(t)\delta_{1i}\,,\label{eq:mom}
\end{equation}
non-dimensionalised by the bulk velocity $U_b$, the half-height of the channel $h$ and the total viscosity $\nu_s$, so the Reynolds number is defined as $Re:=U_bh/\nu_s$. The material derivative is defined as $D_t:=\pder{t}{}+u_k\pder{k}{}$. The parameter $\beta$ is the ratio of solvent viscosity to the zero-shear viscosity of the polymer solution. The force term $f(t)$ drives the flow by enforcing constant mass flow, which is applied in the $x$ direction as indicated by the Kronecker tensor $\delta_{ij}$. The polymer stress tensor $T_{ij}$ is obtained from the conformation tensor $C_{ij}=\left\langle q_iq_j\right\rangle$, which represents the local phase average of the product of the end-to-end vector $q_i$ of each polymer molecule. Its transport equation is
\begin{equation}
    D_t{C_{ij}}=C_{ik}\,\pder{k}{u_{j}}+\pder{k}{u_i}\,C_{kj}-T_{ij}
    +\frac{1}{ReSc}\pdder{k}{k}{C_{ij}}\label{eq:C}
\end{equation}
with 
\begin{equation}
    T_{ij}=\frac{1}{Wi}\left(\frac{C_{ij}}{1-C_{kk}/L^2}-\delta_{ij}\right)\label{eq:T}\,.
\end{equation}

The first two terms on the RHS of Eq.~\eqref{eq:C} represent the stretching caused by the flow and the third term (Eq.~\ref{eq:T}) is the entropic term or spring term that tends to bring stretched polymers back to their least energetic configuration of being coiled. Three parameters define the FENE-P model: $L$ the maximum extensibility of polymers, $Wi$ the ratio of polymers' relaxation time scale $\lambda_p$ to the flow time scale $\lambda_f$, and $\beta$ the ratio of the solvent viscosity to the zero shear rate viscosity of the solution. Throughout this article, the flow time scales used to normalize the relaxation time scale is the integral time scale of the flow $h/U_b$. \YD{ Global artificial diffusion is employed to regularize the hyperbolic Eq.~(\ref{eq:C}) \cite{Purnode1996111} with a high Schmidt number $Sc$ of 1000 for all productions runs, consistent with our earlier paper \cite{page2020traveling}.}

\YD{Production runs use $512n_x\times513$ grids and grid convergence is verified on $1024n_x\times1025$ for at least one simulation for each regime identified here. Similarly, one simulation per regime is carried out with lower Schmidt numbers to confirm that the regime is not an artefact of energy build up at small scales. The grid is stretched in the wall-normal direction and the height of the first cell is critical to resolution of wall polymer stress in chaotic regimes. For the Reynolds number considered, the appropriate wall resolution is found to be $\Delta z_{min}/h=10^{-4}$, which is about two orders of magnitude smaller than Kolmogorov scale. The time step is $10^{-3}h/U_b$ for most simulations and as low as $10^{-4}h/U_b$ for simulations with large $L$ and $Wi$ for numerical stability reasons. }

\YD{Lastly the impact of the Schmidt number is assessed for at least one simulation per regime of EIT. As the Schmidt number increases, the number of points where $C_{ij}$ loses its positiveness increases. This observation must be weighted against the destruction of small scale dynamics for small Schmidt numbers. More information on the effects of the Schimdt number can be found in Appendix. It is important to note that (i) the loss of positive definitiveness of $C_{ij}$ happens mostly in regions where $C_{kk}$ is small and (ii) reducing the time step appears to eradicate the problem, at the expense of increasing the computational cost by two or three orders of magnitude, without changing the dynamics.}

\YD{Supporting materials for the grid convergence study are provided in the Appendix.}

\section{Visual identification of the different structures}
\begin{figure}[t]

\centerline{\includegraphics[width=
\columnwidth]{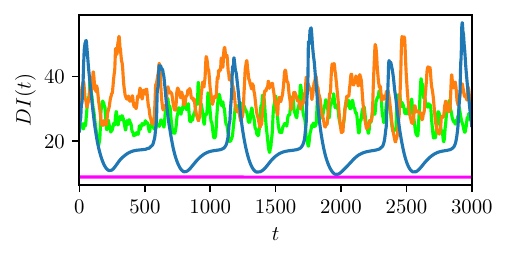}}

\caption{\label{fig:2} Temporal signals of drag increase $DI$ for examples of the four different regimes highlighted in \figref{fig:1}. Colors correspond to simulations introduced in Fig. 1.}
\end{figure}
The identification of the different regimes begins in a domain of fixed length, $L_x=2\pi$ (the influence of the domain's length will be discussed later).
At $Re=1000$, the investigation of flows with $L=50,100,200,500$, $0\leq Wi\leq 250$ and $\beta=0.9$, as well as $\beta=0.97$ and 0.5 (neither shown here) reveals four distinct regimes based on the spatial and temporal structure of polymer stress and pressure. The polymer stress is represented by the first normal stress difference defined as 
\begin{equation}
    N_1:=\widetilde{T}_1-\widetilde{T}_2 \,,\label{eq:N1}
\end{equation} 
where $\widetilde{T}_i$ are the eigenvalues of $T_{ij}$. The choice of $N_1$ is motivated by its more obvious physical role in the well-known rod-climbing and die-swell phenomena \cite{bird1987dynamics}. Note that the structures captured by contours of $N_1$ are similar to structures of the elastic energy
\begin{equation}
e_p:=-\frac{1}{2}\frac{1-\beta}{ReWi}L^2\ln ({1-C_{kk}/L^2})\,. \label{eq:ep}
\end{equation}
Figure~\ref{fig:1} shows snapshots of contours of $N_1$ and $p$ that are representative of the different regimes observed in the $(n_x=1)$-domain across our parameter space. 

The first regime, labelled the chaotic regime (CR), is the original state in which EIT was discovered \cite {dubief2010polymer,dubief2012analysis,Samanta2013el,Dubief2013hh} and observed in subsequent studies \cite{sid2018two,ashwin2019critical}. CR consists of thin sheets of large polymer stress emanating from the near-wall region and stretching toward the centerline at a shallow angle (\figref{fig:1}a). The pressure signature (\figref{fig:1}b) is reminiscent of that corresponding to Tollmien-Schlichting (TS) waves, a series of alternating low and high pressure regions, whose vertical extent does not exceed the channel half-height. This comparison is in agreement with the instability discussed by Ashwin~\textit{et al.} \cite{ashwin2019critical}. 

In Figs.~\ref{fig:1}(a,b), regions of low- and high-pressure in the lower half of the channel are identified by green dots to search for a correlation between the dynamics of $p$ and $N_1$. When the undulations of a sheet of large $N_1$ are locally convex (concave) with respect to their distance from the wall, the local pressure is high (low). The same observation can be made for the top half of the channel. Undulations of high $N_1$ or polymer stress sheets are a signature of chaos in EIT.

Figures~\ref{fig:1}(c,d) reveal the emergence of a peculiar structure: the arrowhead structure. In this second regime, labelled the chaotic arrowhead regime (CAR), two sheets of large $N_1$, one from the upper half, the other from the lower half, join at the centerline. The near-wall structure of sheets is similar to that of CR. The pressure field (\figref{fig:1}d) appears to be dominated by the arrowhead with a low-pressure region ahead of the junction and a high-pressure region in the wake of the junction. This regime remains very chaotic, however the arrowhead remains a robust structure, observable throughout the whole duration of the simulation $4000 h/U_b$. 

The third regime, labelled the intermittent arrowhead regime (IAR), undergoes extended periods of quasi-steadiness, and short periods of chaos. Figures~\ref{fig:1}(e,f) depict an instant of intense chaos, whereas periods of quasi-steadiness are similar to the fourth regime discussed below (Figs.~\ref{fig:1}g,h). The striking difference between CAR and IAR is the higher definition of the arrowhead structure and the presence of a bullet-shaped high pressure region, in the wake of the junction. High- and low-pressure regions near the wall follow the same correlation with the local undulations of $N_1$-sheets as observed in CR.

The fourth regime, labelled the steady arrowhead regime (SAR), is the the main discovery of this article. The structure, shaped like an arrowhead, exhibits perfect symmetry across the centerline as depicted in Figs.~\ref{fig:1}(g,h). The pressure distribution (\figref{fig:1}b) shows a bullet-shaped high-pressure region located in the inside of the arrowhead, and shock-like feature perpendicular with the flow located at short distance downstream of the nose.  Away from the shock, both upstream and in the wake of the arrowhead's junction,  isobars are perpendicular with the flow, as one would expect for a laminar flow. \YD{The convection speed was measured with two approaches. The first is traditional space-time correlations. The second uses an optimization method (standard steepest gradient) to minimize the cost function $\min_{U_C}\left\Vert N_1(x,y,n\Delta T_s)-N_1(x+U_c\Delta T_s,y,(n+1)\Delta T_s)\right\Vert_\infty$ where $\Delta T_s=1 h/U_b$ is the sampling frequency of full flow and polymer fields and $U_c$ is the solution of the optimization process, the convection speed. 
}

\YD{The supplemental material (SI) shows 4 sample movies of the 4 cases displayed in Fig.~\ref{fig:1}. The movie corresponding to SAR shows the same flow twice. The upper image is $N_1(x,y,t)$ and the lower is $N_1(x+U_c(t-t_0),y,t)$, where $t_0$ is the time of the first frame. The authors recognize that the transitions between regimes merit careful attention as their study is likely to hold critical clues to mechanisms creating chaos in EIT flows. Such a study however deserves its own publication.}

\section{Regimes and drag increase}
From \figref{fig:1}, the distinction between CAR and IAR is not obvious. These regimes find their names in the temporal signal of drag increase (DI), defined as the percentage increase relative to the laminar drag for the same viscoelastic conditions. In a constant mass flow simulation, the drag is proportional to the pressure gradient necessary to impose the prescribed mass flow. As depicted in \figref{fig:2}, the temporal evolution of the CR of Figs.~\ref{fig:1}(a,b) is highly disordered or chaotic. The simulation corresponding to the CAR is also chaotic. The intermittent regime (IAR) is quasi periodic, with large fluctuations of $DI$. The large peaks of $DI$ corresponds to the type of chaotic state displayed in \figref{fig:1}(e), the valleys to a quasi-steady arrowhead structure. Finally the signal for SAR is flat 
as it should be for a travelling wave.

\begin{figure}[t]

\centerline{\includegraphics[width=
\columnwidth]{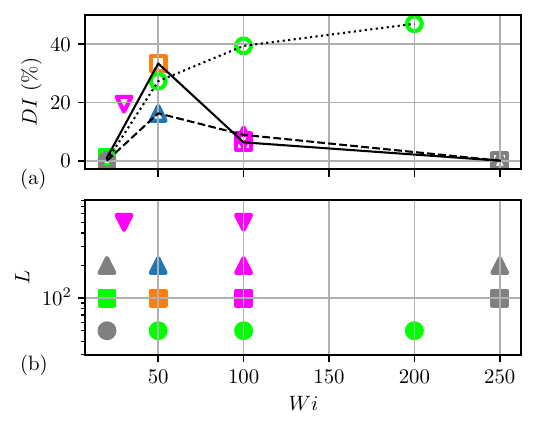}}

\caption{\label{fig:3} (a) Drag increase for different $L$ and $Wi$, $\beta=0.9$, $Re=1000$ and a domain length of $L_x=2\pi$.
\mycircle{black}, $L=50$; 
\mysquare{black}, $L=100$; 
\mytriangle{black}, $L=200$;
\myinvtriangle{black}, $L=500$.
Symbols are color-coded by states as defined in \figref{fig:1} and grey defines the laminar regime. (b) Flow regimes in the $Wi$-$L$ phase space. $Re_b=1000$ and $\beta=0.9$.}
\end{figure}

\begin{figure}[t]
\centerline{\includegraphics[width=
\columnwidth]{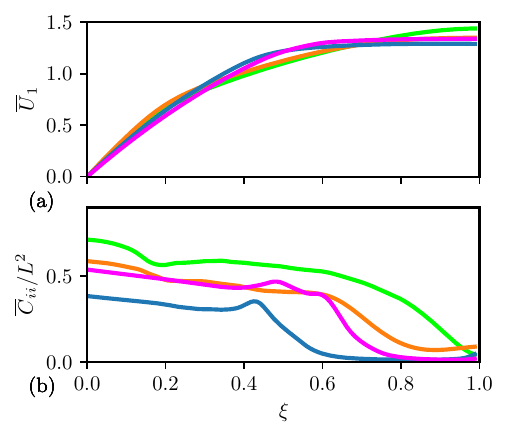}}
\caption{\label{fig:4} (a) Profiles of mean streamwise velocity as a function of the distance from the wall $\xi=1-z$. (b) Profiles of mean trace of the configuration tensor normalized by the polymer maximum extensibility as a function of $\xi$. Colors correspond to simulations introduced in \figref{fig:1}.}
\end{figure}

The average of the temporal signal of $DI$ is performed on several thousands of characteristic time scale $h/U_b$ ranging from 4000 for CR to 10,000 for one of our SAR simulations (6,000 for the others). Technically, the long integration time is not necessary for statistical convergence, but it confirms the steadiness of the solution. The more chaotic the regime is, the faster statistical convergence is achieved. Figure~\ref{fig:3}(a) maps the  drag increase as a function of the Weissenberg number $Wi$ and the polymer maximum extensibility $L$ for the $L_x=2\pi$-domain. Experiments in channel flows \cite{varshney2018drag}  report a non-monotonic behavior of $DI$ as a function of the (non-dimensional) relaxation time $Wi$, where $DI$ increases from $Wi=0$ to a maximum followed by a gradual return to zero (laminar flow)  at high $Wi$. Laminarization was also observed in pipe flows \cite{choueiri2018exceeding}. Figure~\ref{fig:3}(a) suggests a similar behavior  for $L=100$ and 200. Yet we cannot conclusively establish that the dynamics of our simulations leading to the DI evolution is the same as  experiments. As it will be shown below, the length of the domain plays a significant role in the type of regime that may exist at a given combination $(Wi,L)$. For $L=50$, the existence of a maximum of $DI$ could not be established because low-$L$ and high-$Wi$ flows require much lower time steps for stability reasons. Attempts to simulate $L=50$, $\beta=0.9$, and $Wi>200$ proved to be too computationally expensive and too unstable for the present algorithm. This has shown to be reliable and robust for large $L$, $Wi$ and $\beta$, however the present study explores flows where $C_{ii}$ intermittently reaches values very close to $L^2$, which can produce numerical instabilities. A new algorithm is in development to address this issue. 
 
Figure~\ref{fig:3}(b) provides a rough outline of the different regimes in the $Wi$-$L$ phase space for $Re = 1000$, $\beta=0.9$ and a domain length of $L_x=2\pi$. CR is confined to $L=50$ with the exception of $L=100$, $Wi=20$.  For $L=100$, increasing $Wi$ yields an evolution from CR to CAR to SAR to laminar, whereas $L=200$ undergoes an evolution from laminar to IAR to SAR to laminar. For $L=500$, simulations at $Wi=30,100$ reached SAR. 
 
 Mean velocity profiles (\figref{fig:4}a) evolve from a nearly parabolic velocity profiles at CR and CAR to more of a plug flow-like profile at IAR and SAR, with a velocity plateau extending over $\xi\gtrsim 0.5$ 
 These profiles correspond to the flows depicted in \figref{fig:1}. Figure~\ref{fig:4}(b) is a measure of the mean polymer extension throughout the channel. In the CR-flow, the polymer extension is large, above 50\% of $L^2$ over 75\% of the channel half-height. A common characteristic of CR and CAR is the inflexion point around $\xi\approx 0.2$. For IAR and SAR, the polymer extension decreases linearly up to $\xi\approx 0.4$, where it experiences a local maximum before decreasing rapidly to very small values in the core. 
 
 \section{Energy transfers}
%
%
\begin{figure}[t]
\centerline{\includegraphics[width=
\columnwidth]{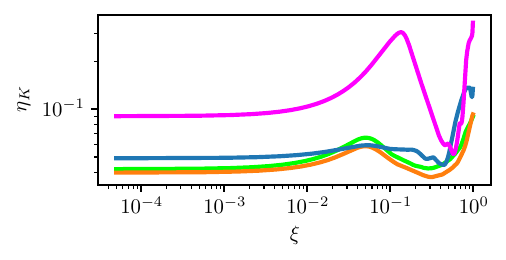}}
\caption{\label{fig:5} Profiles of the mean Kolmogorov scale (Eq.~\ref{eq:etaK}) as a function of the distance from the wall $\xi=1-z$. Colors correspond to simulations introduced in \figref{fig:1}.}
\end{figure}

\begin{figure}[t]
\centerline{\includegraphics[width=
\columnwidth]{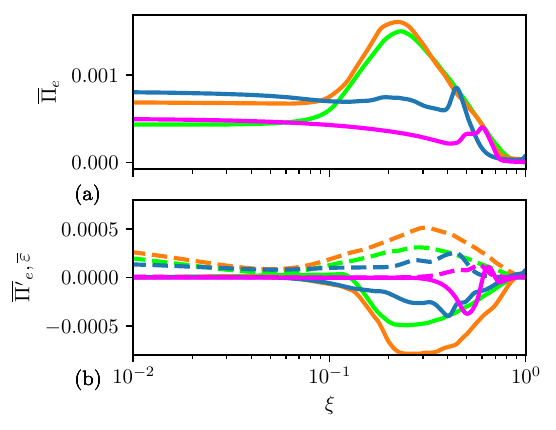}}
\caption{\label{fig:6} Profiles of the mean energy transfer term $\Pi_e$ (a), and, in (b), the mean fluctuating energy transfer term $\overline{\Pi'}_e$ (solid lines) and dissipation rate of TKE $\bar{\varepsilon}$ (dashed lines) as a function of the distance from the wall $\xi=1-z$. Colors correspond to simulations introduced in \figref{fig:1}.}
\end{figure}

 \begin{figure}[t]
\centerline{\includegraphics[width=
\columnwidth]{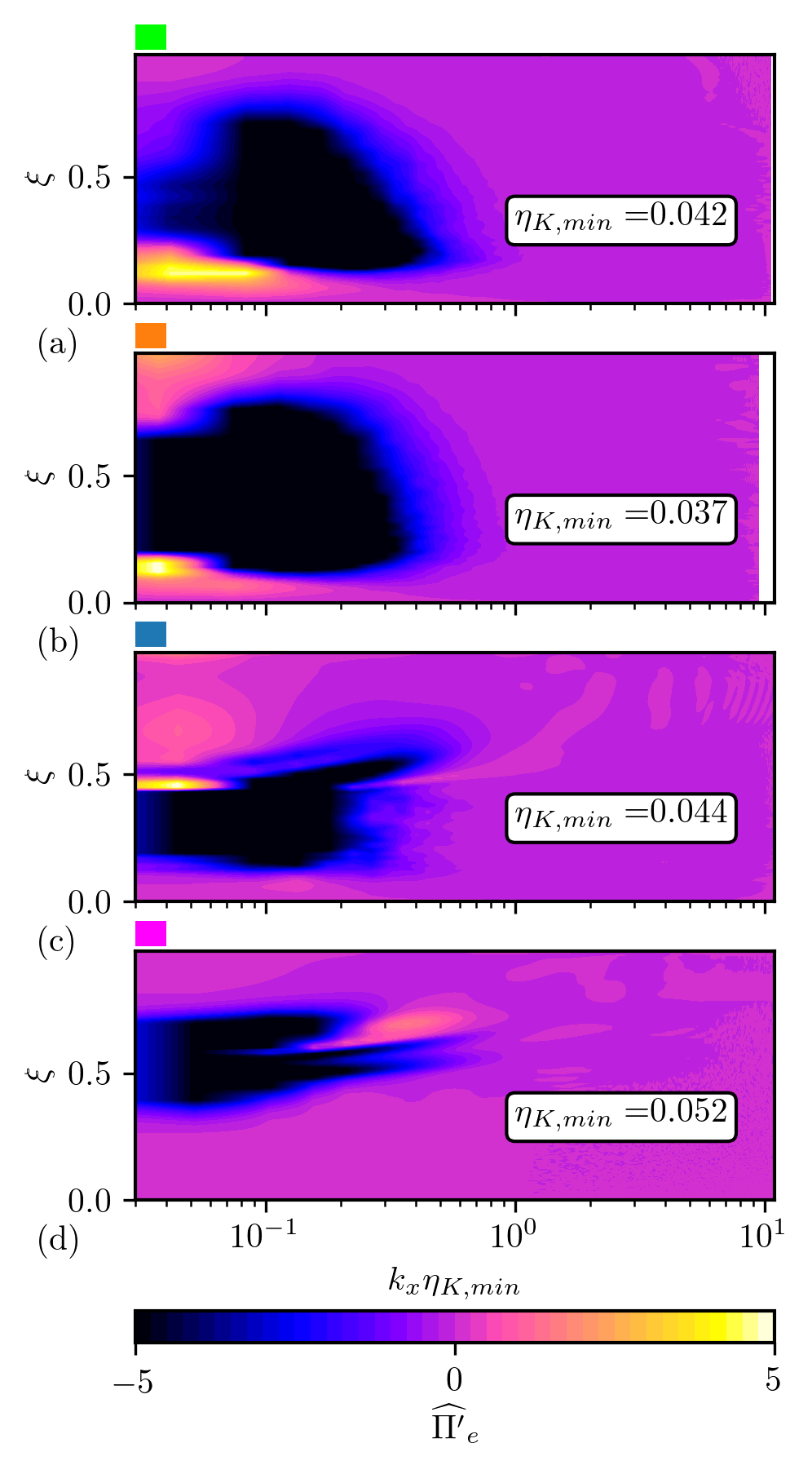}}
\caption{\label{fig:7} Streamwise cospectra of the fluctuations of the energy transfer term defined in Eq.~(\ref{eq:Piehat}) as a function of the distance from the wall $\xi$. The streamwise wavenumber is normalized by the minimum mean Kolmogorov length scale from \figref{fig:3}. (a): CR; (b): CAR; (c): IAR; (d): SAR. Each graph corresponds to simulations introduced in \figref{fig:1}. }
\end{figure}
 
 The transport equations of the  kinetic energy  (KE) $e_u:=\tfrac{1}{2}u_iu_i$ and elastic energy (EE) $e_p$ (defined in Eq.~\ref{eq:ep}) read 
 \begin{equation}
    \begin{split}
        \pder{t}{e_u}+u_k\pder{k}{e_u}=&-\pder{i}{u_ip}+\frac{\beta}{Re}\pdder{k}{k}{e_u}-\frac{\beta}{Re}\left(\pder{k}{u_i}\right)\left(\pder{k}{u_i}\right)\\
        &+\frac{1-\beta}{Re}\pder{k}{(u_kT_{ik})}-\Pi_e\label{eq:Te_u}
    \end{split}
\end{equation}
and
 \begin{equation}
    \pder{t}{e_p}+u_k\pder{k}{e_p}=-\frac{1}{2}\frac{1-\beta}{ReWi}fT_{ii}+\Pi_e\label{eq:Te_p} \,,
\end{equation}
where
\begin{equation}
    \Pi_e:=\frac{1-\beta}{Re}T_{ij}S_{ij}\label{eq:Pie}
\end{equation}
is the energy transfer between KE and EE. Similar transport equations can be derived for the turbulent kinetic energy (TKE, $e'_u:=\tfrac{1}{2}u'_iu'_i$) and turbulent elastic energy (TEE, $e'_p$), yielding
 \begin{equation}
    \Pi'_e:=\frac{1-\beta}{Re}T'_{ij}S'_{ij}\label{eq:Piep}.
\end{equation}
Also of interest to our study, the dissipation rate of TKE, \begin{equation}
    \varepsilon:=\tfrac{\beta}{Re}\pder{j}{u'_i}\pder{j}{u'_i}\,, \label{eq:eps}
\end{equation}
defines the Kolmogorov length scale, here written with our adopted normalization, 
\begin{equation}
    \eta_K := \left(\frac{(\beta/Re)^3}{\bar{\varepsilon}}\right)^\frac{1}{4}\label{eq:etaK}\,,
\end{equation}
which is the smallest scale, or dissipation scale, in classical turbulence. The length scales in EIT are yet to be defined, since they most likely depend upon flow and polymer parameters at a minimum. For now, the minimum mean Kolmogorov length scale over the height of the channel is adopted as a reference length scale for the flow. The distribution of the Kolmogorov length scale for the four different regimes and simulations of \figref{fig:1} are shown in \figref{fig:5}. The smallest length scale is at a distance from the wall $\xi$ ranging from 0.3 for the chaotic regimes to 0.6 for SAR. CR, CAR and SAR show a local maximum in the range $0.05\lesssim\xi\lesssim0.2$ and another at the centerline. The latter maximum is not surprising since all regimes experience little to no polymer stress with the exception of the junction of the arrowhead. 

Figure~\ref{fig:6} shows the profiles of the mean and fluctuating energy transfer terms throughout the half-height of the channel. From Eqs.~(\ref{eq:Te_u}) and (\ref{eq:Te_p}), positive $\Pi_e$ or $\Pi'_e$ indicates an energy transfer from the mean kinetic energy of the flow to the mean elastic energy or from TKE to TEE, respectively. The mean energy transfer is positive throughout the channel (\figref{fig:6}a), with a maximum at CR and CAR around $\xi\approx 0.2$, where profiles of $C_{ii}/L^2$ (\figref{fig:4}b) show an inflection point. For IAR and SAR, the local maximum corresponds to the steep decrease in $C_{ii}/L^2$.  

The fluctuating energy transfer is negative for CR, CAR and IAR, showing that energy is flowing from TEE to TKE on average (\figref{fig:6}b). For SAR, the energy transfer $\overline{\Pi'}_e$ switches sign close to the centerline. The very near wall region is void of fluctuating energy transfer up to $\xi\approx 0.1$. This region extends further for SAR, to $\xi \approx 0.3$. Naturally for SAR, fluctuations are in fact spatial, since the flow is invariant by translation, which explains why the dissipation rate of TKE is null in the very near-wall region, whereas $\bar{\varepsilon}$ for other regimes is finite. 

Figure~\ref{fig:7} investigates the spectral representation $\widehat{\Pi'}_e$ of the energy transfer
 \begin{equation}
     \widehat{\Pi'}_e:=\frac{1-\beta}{Re}\widehat{T}_{ij}*\widehat{S}_{ij}\,,\label{eq:Piehat}
 \end{equation}
 which is proportional to the co-spectra of $T'_{ij}S'_{ij}$. In Eq.~(\ref{eq:Piehat}) the $\widehat{a}$ symbol defines the Fourier transform of variable $a$. This analysis allows for the investigation of the scales and distances from the wall at which polymers gain energy from the flow and \textit{vice versa}. The streamwise wavenumber is normalized by the minimum of the Kolmogorov length scale. 
 
 For CR (\figref{fig:7}a), the energy transfer from TEE to TKE\YD{, \textit{i.e.} $\widehat{\Pi}_e$,} is mostly from $10\eta_{K,min}$ down to $\sim 5\eta_{K,min}$ \YD{at a distance from the wall $0.2\lesssim\xi\lesssim 0.75$}. The energy transfer from TKE to TEE occurs at larger scales $\gtrsim10\eta_{K,min}$ \YD{but closer to the wall, $\xi\sim0.1-0.2$}. EIT in its chaotic regime is therefore sustained by an upscale energy transfer from polymers to flow and a downscale energy transfer from flow to polymers. \YD{The physical location of the latter corresponds to} the location of the inflection point of $C_{ii}/L^2$ (\figref{fig:4}b) and the maximum mean transfer of energy from KE to EE (\figref{fig:6}a). The upscale energy transfer dominates the fluctuating energy transfer ($\overline{\Pi'}_e$) as shown in \figref{fig:6}b. 
 The flow dynamics of CR-EIT stretches polymers to large mean extension levels. Stretched polymers are organized in thin-sheets, which in turn feed TKE via a mechanism yet to be identified, but occurring at smaller scales than the mechanism of polymer stretching.
 
 Whereas the picture of energy transfer at CAR (\figref{fig:7}b) is similar to CR, IAR and SAR show a much different pattern (Figs.~\ref{fig:7}c,d). The energy transfer from TKE to TEE shifts upward and is still large scale for IAR but becomes small scale for SAR, albeit of weaker intensity than for all other regimes. The region of energy transfer from TEE to TKE is truncated above $\xi\approx0.5$  for IAR and again shifted upward and narrower for SAR. A major distinction between chaotic and steady or quasi-steady regimes may be in the location of the energy transfer from TKE to TEE ($\widehat{\Pi'}_e>0$). One could speculate that the origin of chaos is not only in the spectral locality of the energy transfers but also spatial locality. Injection of TKE into TEE in the region where polymers are the most stretched (near-wall region) could conceivably excite instabilities in the sheets of high polymer stress resulting into the undulations observed in \figref{fig:1}(a) correlated to regions of high- and low-pressure \figref{fig:1}(b).

 \section{Influence of domain length, Reynolds number, and $\beta$}
 
 \begin{figure}[t]
\centerline{\includegraphics[width=\columnwidth]{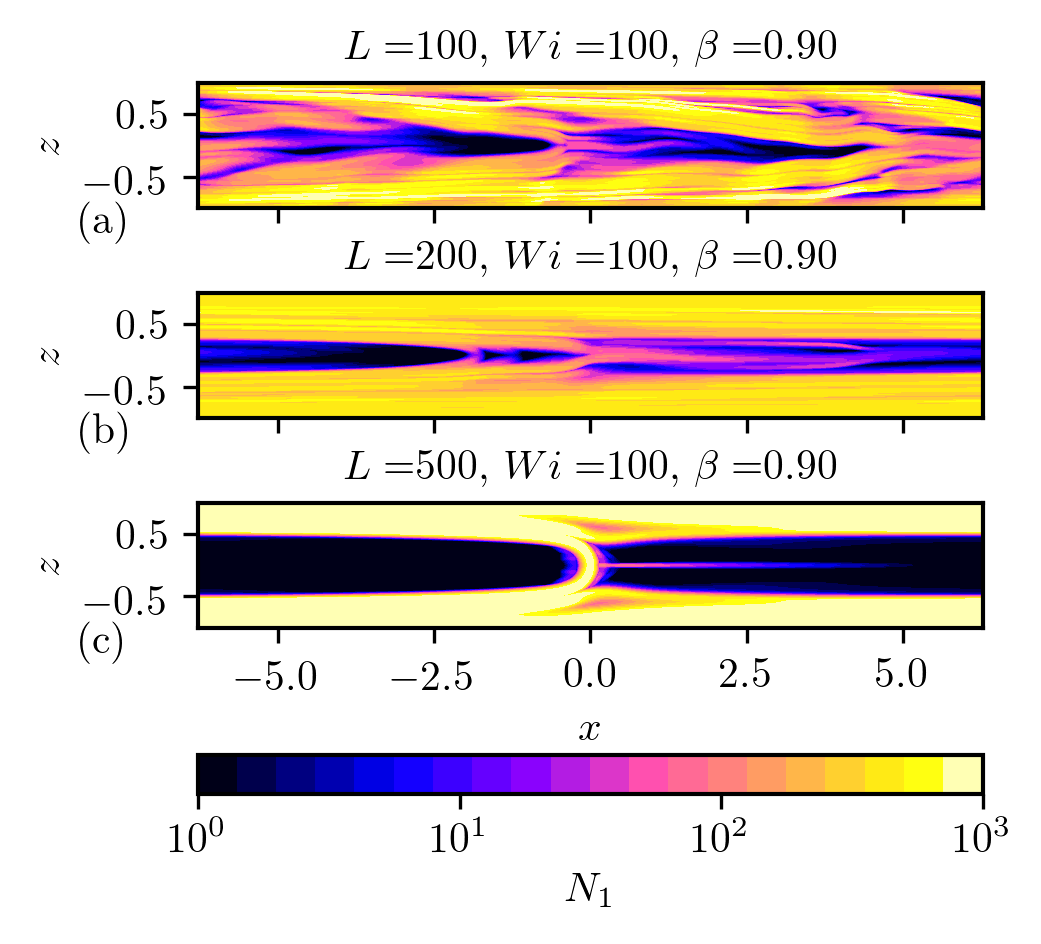}}
\caption{\label{fig:8} Snapshots of $N_1$ fields in a domain twice the length of the domain used in simulations used for Fig.  \ref{fig:1}. For all simulations, the Reynolds number is 1000.}
\end{figure}

For the chaotic regime, we verified that statistics are not affected by doubling the length of the domain (not shown). SAR flows with $L=100$ and 200 prove to be highly sensitive to the domain length. Doubling the domain size with $L=100$, $Wi=100$, $\beta=0.9$ causes the flow to shift from SAR to CAR as shown in \figref{fig:8}(a). The SAR for $L=200$ depicted in \figref{fig:1}(g) becomes IAR in the larger domain (\figref{fig:8}b). SAR is recovered for $L=500$ for $L_x=4\pi$ (\figref{fig:8}a) and $8\pi$ (not shown).

The mechanism driving chaos appears to be a function of both $L$ and $L_x$. In other words, the undulation of sheets of large $N_1$ or polymer stress may be created by a large scale instability that is damped when the domain is too short. In spite of the large number of simulations performed for this article, this could not be ascertain. Identifying the exact wavelength of this possible instability as a function of $L$ requires further simulations to probe the range $L_x\in[2\pi,4\pi]$. 
The extensibility parameter $L$ drives the length scale of the sheets of large polymer stress, as well as the intensity of the first normal stress difference, which both increases with $L$ for a given $Wi$. 

 \begin{figure}[t]
\centerline{\includegraphics[width=\columnwidth]{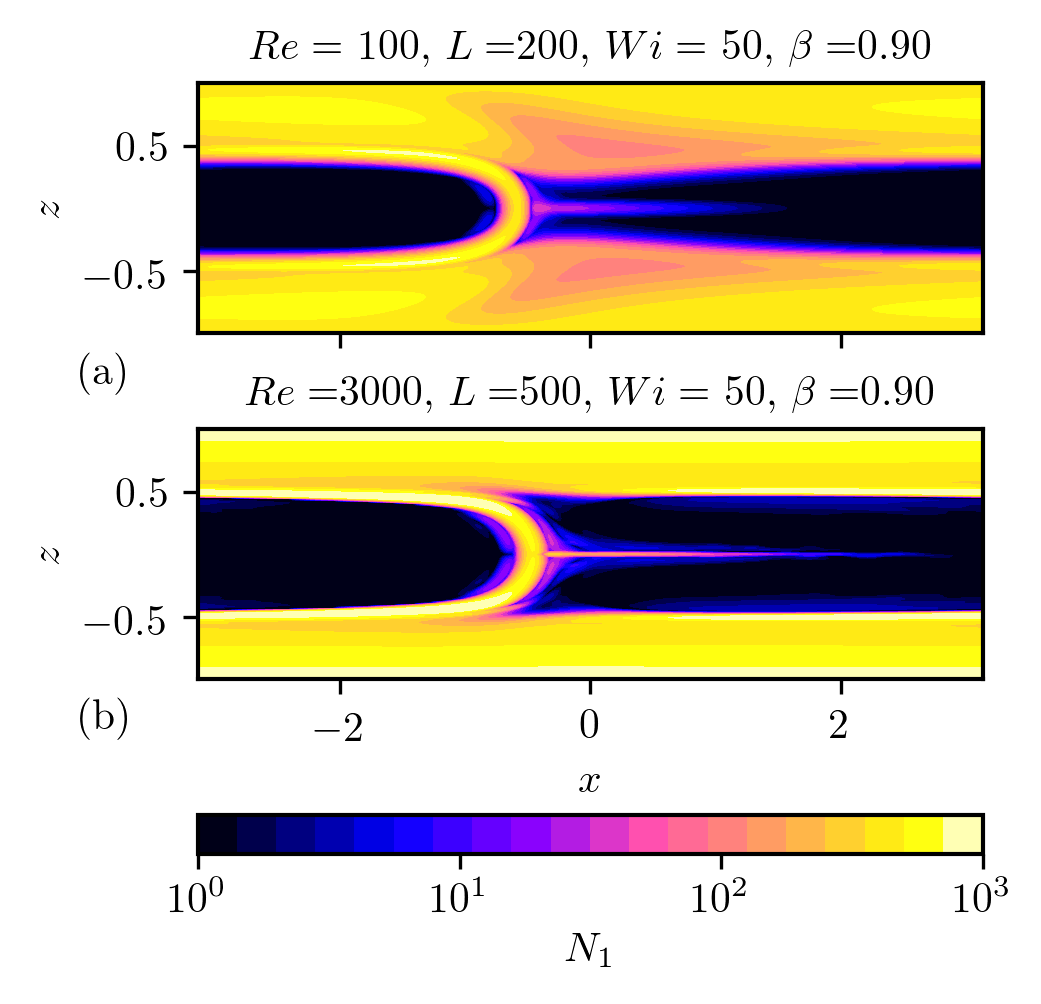}}
\caption{\label{fig:9} Snapshots of $N_1$ fields of steady arrowhead regime at two different Reynolds numbers.}
\end{figure}
The arrowhead structure, in its steady form, is found to exist at least between $Re=100$ and $3000$ (Figs.~\ref{fig:9}a,b, respectively). At $Re=100$, a depression in the first normal stress difference is observed in the near-wall region at the front of the arrowhead's junction, similar to the one observed in \figref{fig:1}(g) at $Re=1000$, but extending almost to the wall and with a smaller gradient with the surrounding stress. Note that at $Re= 3000$ SAR requires longer the extensibility parameter $L$. There was no attempt to reduce the length of the computational domain to investigate whether IAR and CAR could become stable for lower $L$ at this Reynolds number. 

 \begin{figure}[t]
\centerline{\includegraphics[width=\columnwidth]{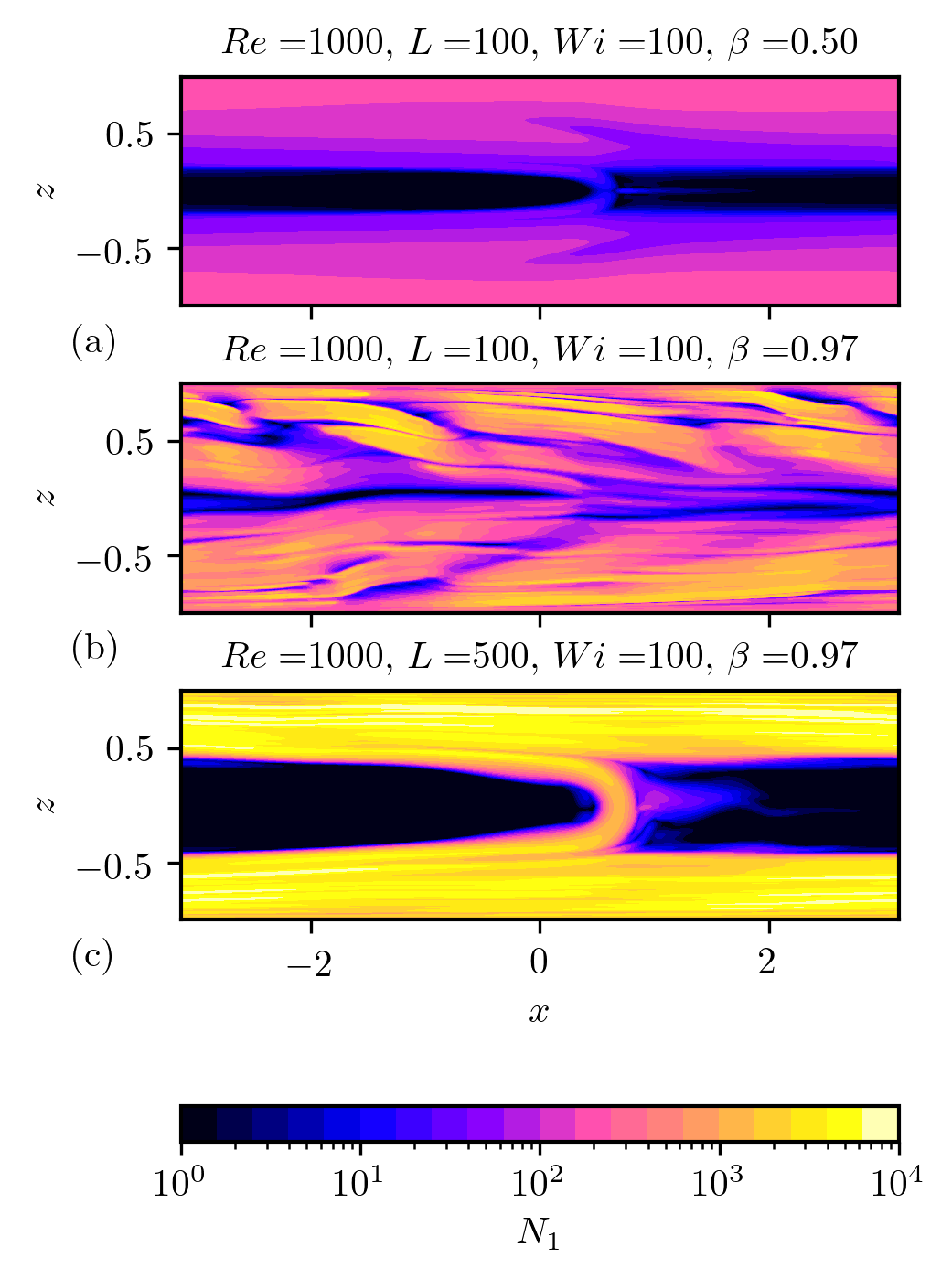}}
\caption{\label{fig:10} Snapshots of $N_1$ fields illustrating the effect of the parameter $\beta$. (a) and (b) share the same $L=100$ and $Wi=100$ as SAR simulations identified for $\beta = 0.9$ (see Fig. \ref{fig:3}b). $\beta=0.5$ sustains SAR (a), whereas $\beta=0.97$ destabilizes the flow to CR. (c) shows SAR for $\beta=0.97$ at $L=500$ and $Wi=100$.}
\end{figure}

Lastly, \figref{fig:10} illustrates the effect of the ratio $\beta$  first on two different simulations at $Re=1000$ with $L=100$ and $Wi=100$ in a $L_x=2\pi$-domain (Figs~\ref{fig:10}a-b). At $\beta=0.9$, the flow achieves SAR (see Fig. \ref{fig:3}b). The lower $\beta=0.5$ simulation shows the same regime, confirmed over 4,000 $h/U_b$. There are some visible differences in the shape and width of the arrowhead but the main features of the arrowhead structure remain clearly identifiable.  Increasing $\beta$ to 0.97 triggers the chaotic regime CR, as shown in Fig. \ref{fig:10}b. Keeping $Wi=100$ and $\beta=0.97$, SAR can be recovered at $L=500$ as shown by Fig.~\ref{fig:10}c. We have not sought to define precisely the critical $L$ at which the flow is stabilized, nor the critical $\beta\in[0.9,0.97]$ at which the flow transitions from SAR to intermittence or chaos. A future investigation of the influence of $\beta$ is however necessary and will be conducted in the near future. The data shown here, suggests two possible co-existing polymer effects. Shear thinning, driven by low $\beta$ may help stabilize the very near region where the stretch is the highest. At the other end of the spectrum, when $(1-\beta)$ approaches zero, the recovery of SAR might be indicative of the role of extensional viscosity in the structure of the arrowhead. Tamano \textit{et al.}\cite{tamano2009effect} showed some similarity in the drag reducing properties of flows with comparable $1-\beta)L^2$. Noticeably the most significant drag reduction obtained by Tamano \textit{et al.} was for \YD{$(1-\beta)L^2=10^3$}, the highest value achieved in their simulations. In the limited data for the present study, SAR is also observed for solution with $(1-\beta)L^2\ge 10^3$, the simulation of SAR shown in Figs~\ref{fig:10}a and c have $(1-\beta)L^2=5\times 10^3$ and $7.5\times 10^3$, respectively, whereas the CR regime shown in \figref{fig:10}b is at $(1-\beta)L^2=3\times 10^2$.

\section{Steady arrowhead regime}
 \begin{figure}[t]
\centerline{\includegraphics[width=\columnwidth]{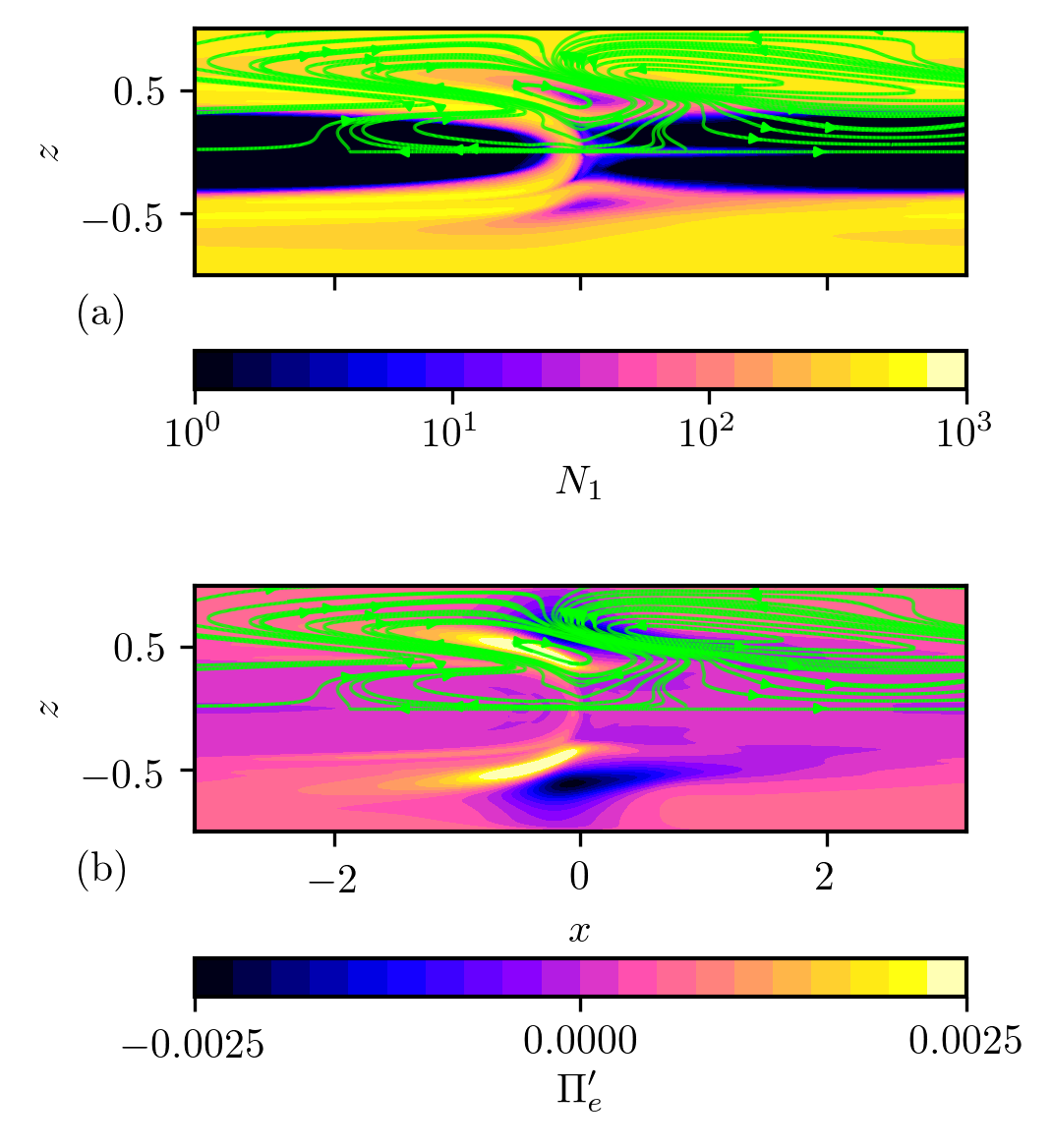}}
\caption{\label{fig:11} (a) Superimposition of streamlines computed from the fluctuating velocity field and contours of $N_1$ for $Re=1000$, $L=200$, $Wi=100$ and $\beta=0.9$ (\figref{fig:1}g). (b) Superimposition of the same streamlines and contours of the transfer of energy fluctuations $\Pi'_e$}
\end{figure}

The steady arrowhead regime proves to be a robust feature of EIT. It appears to be triggered by the ratio of polymer extensibility to domain streamwise length $L_x$ for polymer extensibility larger than a threshold. This threshold $L_{SAR}$ is in the range $L_{SAR}\in]50,100]$ for $\beta=0.9$ and $Re=1000$ according to our data. Further numerical experiments are needed to understand the relationship between $L$ and $L_x$ and its influence on chaos. Regarding the latter point, one may speculate that the intense high-pressure region in the wake of the arrowhead's junction stabilizes the flow. When $L_x$ is increased, chaos may arise in regions that are far enough away from the junction, possibly through a linear instability \cite{garg2018viscoelastic,chaudhary2020linear}. Under this scenario, the intensity of the pressure gradient between the front and back of the junction may be the determinant factor.

Figure~\ref{fig:11} illustrates the complexity of SAR, through streamlines of the fluctuating velocity field superimposed on contours of $N_1$ (\figref{fig:11}a) and superimposed on contours of the transfer of energy fluctuations $\Pi'_e$. Streamlines show the existence of two large scale structures in the near wall regions whose interface is located in a region where energy is transferred from TKE to TEE (\YD{$\Pi'_e>0$}). Unsurprisingly, the snapshot of $\Pi'_e$ in physical space resembles the energy transfer in the spectral-distance from the wall space (\figref{fig:7}d). Figure~\ref{fig:11} also establishes the correlation between the depletion of polymer stress in the upper and lower front of the junction and the interface between energy transfers from TKE to TEE and from TEE to TKE (\YD{$\Pi'_e<0$}). SAR's energy transfers are localized and appear to drive the dramatic perturbations of velocity fluctuations. 

The arrowhead structure most likely owes its symmetry to the presence of walls. A similar structure was observed in a Kolmogorov flow \cite{berti2008two,berti2010elastic} at $Re\lesssim1$, a regime considered to be the upper bound of elastic turbulence. Figures 7b and 8a of Berti \& Boffetta\cite{berti2010elastic} clearly show thin sheets of large $T_{11}$ joining in a pattern similar to the arrowhead. The similarity between elastic turbulence and EIT suggests that the same fundamental mechanisms of polymer/flow interactions may be at play. 

\YD{The robustness of the arrowhead at low Reynolds number was demonstrated in Page \textit{et al.} \cite{page2020traveling}, for $L=500$, down to Re=60 and for relatively small $Wi={\cal O}(10)$.}

\begin{table}[t]
    \centering
    \begin{tabular}{|l|ccc|c|}
    \hline
      Regime & $L$ & $Wi$ & $\beta$   & $\rho_p$  \\
      \hline
        CR & 50 & 50 & 0.9 & 0.23 \\
        CAR & 100 & 50 & 0.9 & 0.48 \\
        IAR & 200 & 50 & 0.9 & 0.9 \\
        SAR & 200 & 100 & 0.9 & 1.0\\
        \hline
    \end{tabular}
    \caption{Caption}
    \label{tab:corrp}
\end{table}
\YD{
\section{Regime identification}}

\YD{
The discovery process detailed so far informs the derivation of a possible identification criterion for the four regimes, CR, CAR, IAR and SAR, specific to 2D simulations of periodic channel flows in a relatively short domain. Although the extension of such criterion to longer domains in 2D and 3D periodic or spatially developing flows is not straightforward, the aim is to use primary flow variables, velocity and pressure, that are accessible experimentally.  The first component of the criterion is based on wall-pressure fluctuations which have been used to characterize ET \cite{jun2009power} and EIT \cite{Samanta2013el,choueiri2021experimental}. The correlation of two wall-pressure signals collected at the same streamwise location but on opposite walls,}
\begin{equation}
    \rho_p = \frac{\overline{p'_{(x,z=+h,t)}p'_{(x,z=-h,t)}}}{\sqrt{\overline{p'^2}_{(z=+h)}\overline{p'^2}_{(z=-h)}}}
    \,,
\label{eq:corrp}
\end{equation}
\YD{is a measure of the symmetry of the flow about the centerline. For Reynolds number $Re_b=1000$, the values of this correlation are reported in Table~\ref{tab:corrp} and can be categorized as low ($\rho_p\sim 0.2$, CR), moderate ($\rho_p\sim0.5$, CAR), high ($\rho_p\sim 0.9$, IAR) and perfect ($\rho_p=1$, SAR). The exact bounds delimiting two adjacent regimes require further investigation of the parameter space $(Re,Wi,L,\beta)$ and will be the focus of future research. The wall-pressure correlation criterion can objectively identify the perfect or near perfect symmetry imposed by the arrowhead structure in IAR and SAR, but remains a necessary condition not a sufficient condition to establish the existence of an arrowhead. 
}

\begin{figure}[t]
\centerline{\includegraphics[width=\columnwidth]{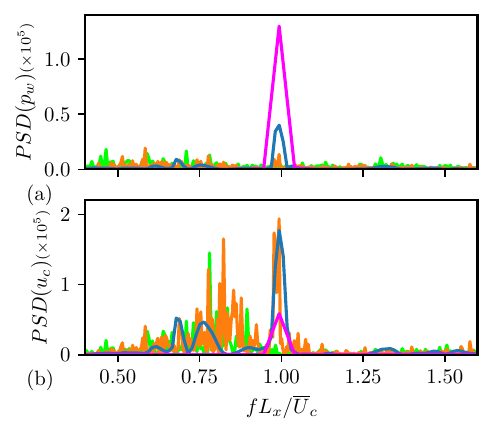}}
\caption{\label{fig:12} Power spectral density plots of times series sampled at a fixed location: (a) wall pressure; (b) centerline streamwise velocity. Colors correspond to simulations introduced in Fig.~\ref{fig:1}
}
\end{figure}

\YD{Plots of power spectral density (PSD) of the wall-pressure signal  (\figref{fig:12}a) show a distinct peak for IAR and SAR at the frequency corresponding to the flow-through time of the arrowhead based on the mean centerline velocity $\overline{U}_c$ and the length of the domain $L_x$. The signature of the arrowhead in CAR is however too small and too close to the energetic contributions of surrounding frequencies in the range 0.5 $\overline{U}_c/L_x$ to 1.5 $\overline{U}_c/L_x$ to be used as an identification criterion. PSD of the streamwise velocity component at the centerline, a quantity that can be measure experimentally, provides an objective identification of the presence of an arrowhead for CAR, with an energy at $\overline{U}_c/L_x$ comparable to that of IAR.
}
\begin{table}[t]
    \centering
    \begin{tabular}{lccc}
    \hline
    Regime & Correlation & \multicolumn{2}{c}{$PSD$ peak  at $f=U_c/L_x$}\\
    \hline
         & $\rho_p$ & $p_w$ & $u_c$ \\
        CR & Low ($\sim0.2$) & No & No\\
        CAR & Moderate ($\sim0.5$) & No & Yes \\
        IAR & High ($\sim 0.9$) & Yes & Yes \\
        SAR & Perfect ($=1$) & Yes & Yes\\
        \hline
    \end{tabular}
    \caption{Characteristics of the different regimes based on the correlation of wall pressure fluctuations at $z=\pm h$, and the presence or absence of a distinct energetic contribution in the PSD of the wall pressure fluctuations and the centeriline velocity fluctuations.}
    \label{tab:criterion}
\end{table}
\YD{Summarized in Table~\ref{tab:criterion}, the four different regimes can be objectively identified, for a periodic channel flow of length $2\pi$, by a combination of correlation and power spectral density analyses of pressure and velocity, two quantities that experiments can measure. The correlation of two wall pressure signals collected at the same streamwise location and on opposite walls measures the symmetry of the flow. Power spectral densities of streamwise centerline velocity fluctuations identify the presence of arrowhead as a peak in energy contribution at a frequency corresponding to the centerline mean velocity and the length of the domain. Another interesting observation is the patch of energetic contribution for the streamwise centerline velocity fluctuations around $fL_x/U_c\sim0.8$, which is only visible for CR and CAR. Whether this patch is a signature of chaos remains an open question that will be addressed in future research.}

\YD{The proposed criterion works for a periodic channel of dimensions comparable to the streamwise scale of the arrowhead. Under this condition, the stability of the IAR and SAR arrowhead produces a dominant frequency. Simulations in longer computational domains presented here suggest that arrowhead structures might be come intermittent, or puff-like.  For future experimental or computational studies in large domains, the frequency criterion might be advantageously replaced by POD or DMD}.


\section{Conclusion}

This paper describes the discovery of the first coherent structure found in elasto-inertial turbulence. This takes the particularly simple form of a travelling wave in which polymer stress sheets which originate near the walls bend to meet at the channel centre to form a symmetric arrowhead structure. Alongside a regime where this arrowhead is an attractor (SAR), our simulations have also established the existence of several other regimes in EIT - chaotic (CR), chaotic arrowhead (CAR) and  intermittent arrowhead (IAR). These regimes are identified by the structure of the polymer stress field, the fluctuations of the drag increase in time and the energy transfer between polymers and flow. The latter clearly demonstrates that the transfer of energy from fluctuations of elastic energy to fluctuations of  turbulent kinetic energy is an upscale mechanism occurring predominantly at small scales away from the walls. Reverse energy transfer occurs at large scales and is a downscale mechanism with the exception of SAR. For the steady arrowhead state, the reverse energy flow is also at small scales but further away from the walls.

The arrowhead coherent structure is a robust state of the flow, which exists over a large range of Reynolds numbers, polymer extensibility, Weissenberg numbers and parameter $\beta$.  This structure should help uncover the fundamental dynamics underpinning EIT and reveal the mechanism of energy transfer between flow and polymers. We hope to report on further progress in this  direction soon.

\begin{acknowledgments}
This research is supported by the National Science Foundation (NSF-CBET-1805636) and the US-Israel Binational Science Foundation \#31747. The opinions, findings, and conclusions or recommendations expressed are those of the authors and do not necessarily reflect the views of the National Science Foundation or of the US-Israel Binational Science Foundation. 
 
\end{acknowledgments}

\bibliography{bib}

\begin{thebibliography}{31}%
\makeatletter
\providecommand \@ifxundefined [1]{%
 \@ifx{#1\undefined}
}%
\providecommand \@ifnum [1]{%
 \ifnum #1\expandafter \@firstoftwo
 \else \expandafter \@secondoftwo
 \fi
}%
\providecommand \@ifx [1]{%
 \ifx #1\expandafter \@firstoftwo
 \else \expandafter \@secondoftwo
 \fi
}%
\providecommand \natexlab [1]{#1}%
\providecommand \enquote  [1]{``#1''}%
\providecommand \bibnamefont  [1]{#1}%
\providecommand \bibfnamefont [1]{#1}%
\providecommand \citenamefont [1]{#1}%
\providecommand \href@noop [0]{\@secondoftwo}%
\providecommand \href [0]{\begingroup \@sanitize@url \@href}%
\providecommand \@href[1]{\@@startlink{#1}\@@href}%
\providecommand \@@href[1]{\endgroup#1\@@endlink}%
\providecommand \@sanitize@url [0]{\catcode `\\12\catcode `\$12\catcode
  `\&12\catcode `\#12\catcode `\^12\catcode `\_12\catcode `\%12\relax}%
\providecommand \@@startlink[1]{}%
\providecommand \@@endlink[0]{}%
\providecommand \url  [0]{\begingroup\@sanitize@url \@url }%
\providecommand \@url [1]{\endgroup\@href {#1}{\urlprefix }}%
\providecommand \urlprefix  [0]{URL }%
\providecommand \Eprint [0]{\href }%
\providecommand \doibase [0]{https://doi.org/}%
\providecommand \selectlanguage [0]{\@gobble}%
\providecommand \bibinfo  [0]{\@secondoftwo}%
\providecommand \bibfield  [0]{\@secondoftwo}%
\providecommand \translation [1]{[#1]}%
\providecommand \BibitemOpen [0]{}%
\providecommand \bibitemStop [0]{}%
\providecommand \bibitemNoStop [0]{.\EOS\space}%
\providecommand \EOS [0]{\spacefactor3000\relax}%
\providecommand \BibitemShut  [1]{\csname bibitem#1\endcsname}%
\let\auto@bib@innerbib\@empty
\bibitem [{\citenamefont {Samanta}\ \emph {et~al.}(2013)\citenamefont
  {Samanta}, \citenamefont {Dubief}, \citenamefont {Holzner}, \citenamefont
  {Sch{\"a}fer}, \citenamefont {Morozov}, \citenamefont {Wagner},\ and\
  \citenamefont {HOF}}]{Samanta2013el}%
  \BibitemOpen
  \bibfield  {author} {\bibinfo {author} {\bibfnamefont {D.}~\bibnamefont
  {Samanta}}, \bibinfo {author} {\bibfnamefont {Y.}~\bibnamefont {Dubief}},
  \bibinfo {author} {\bibfnamefont {M.}~\bibnamefont {Holzner}}, \bibinfo
  {author} {\bibfnamefont {C.}~\bibnamefont {Sch{\"a}fer}}, \bibinfo {author}
  {\bibfnamefont {A.~N.}\ \bibnamefont {Morozov}}, \bibinfo {author}
  {\bibfnamefont {C.}~\bibnamefont {Wagner}},\ and\ \bibinfo {author}
  {\bibfnamefont {B.}~\bibnamefont {HOF}},\ }\bibfield  {title} {\bibinfo
  {title} {{Elasto-inertial turbulence.}},\ }\href@noop {} {\bibfield
  {journal} {\bibinfo  {journal} {Proceedings of the National Academy of
  Sciences}\ }\textbf {\bibinfo {volume} {110}},\ \bibinfo {pages} {10557}
  (\bibinfo {year} {2013})}\BibitemShut {NoStop}%
\bibitem [{\citenamefont {Dubief}\ \emph {et~al.}(2013)\citenamefont {Dubief},
  \citenamefont {Terrapon},\ and\ \citenamefont {Soria}}]{Dubief2013hh}%
  \BibitemOpen
  \bibfield  {author} {\bibinfo {author} {\bibfnamefont {Y.}~\bibnamefont
  {Dubief}}, \bibinfo {author} {\bibfnamefont {V.~E.}\ \bibnamefont
  {Terrapon}},\ and\ \bibinfo {author} {\bibfnamefont {J.}~\bibnamefont
  {Soria}},\ }\bibfield  {title} {\bibinfo {title} {{On the mechanism of
  elasto-inertial turbulence}},\ }\href@noop {} {\bibfield  {journal} {\bibinfo
   {journal} {Physics of Fluids}\ }\textbf {\bibinfo {volume} {25}},\ \bibinfo
  {pages} {110817} (\bibinfo {year} {2013})}\BibitemShut {NoStop}%
\bibitem [{\citenamefont {Wensink}\ \emph {et~al.}(2012)\citenamefont
  {Wensink}, \citenamefont {Dunkel}, \citenamefont {Heidenreich}, \citenamefont
  {Drescher}, \citenamefont {Goldstein}, \citenamefont {Löwen},\ and\
  \citenamefont {Yeomans}}]{wensink2012meso}%
  \BibitemOpen
  \bibfield  {author} {\bibinfo {author} {\bibfnamefont {H.~H.}\ \bibnamefont
  {Wensink}}, \bibinfo {author} {\bibfnamefont {J.}~\bibnamefont {Dunkel}},
  \bibinfo {author} {\bibfnamefont {S.}~\bibnamefont {Heidenreich}}, \bibinfo
  {author} {\bibfnamefont {K.}~\bibnamefont {Drescher}}, \bibinfo {author}
  {\bibfnamefont {R.~E.}\ \bibnamefont {Goldstein}}, \bibinfo {author}
  {\bibfnamefont {H.}~\bibnamefont {Löwen}},\ and\ \bibinfo {author}
  {\bibfnamefont {J.~M.}\ \bibnamefont {Yeomans}},\ }\bibfield  {title}
  {\bibinfo {title} {Meso-scale turbulence in living fluids},\ }\href@noop {}
  {\bibfield  {journal} {\bibinfo  {journal} {Proceedings of the National
  Academy of Sciences}\ }\textbf {\bibinfo {volume} {109}},\ \bibinfo {pages}
  {14308–14313} (\bibinfo {year} {2012})}\BibitemShut {NoStop}%
\bibitem [{\citenamefont {Alert}\ \emph {et~al.}(2020)\citenamefont {Alert},
  \citenamefont {Joanny},\ and\ \citenamefont
  {Casademunt}}]{alert2020universal}%
  \BibitemOpen
  \bibfield  {author} {\bibinfo {author} {\bibfnamefont {R.}~\bibnamefont
  {Alert}}, \bibinfo {author} {\bibfnamefont {J.-F.}\ \bibnamefont {Joanny}},\
  and\ \bibinfo {author} {\bibfnamefont {J.}~\bibnamefont {Casademunt}},\
  }\bibfield  {title} {\bibinfo {title} {Universal scaling of active nematic
  turbulence},\ }\href@noop {} {\bibfield  {journal} {\bibinfo  {journal}
  {Nature Physics}\ ,\ \bibinfo {pages} {1}} (\bibinfo {year}
  {2020})}\BibitemShut {NoStop}%
\bibitem [{\citenamefont {Groisman}\ and\ \citenamefont
  {Steinberg}(2000)}]{groisman2000elastic}%
  \BibitemOpen
  \bibfield  {author} {\bibinfo {author} {\bibfnamefont {A.}~\bibnamefont
  {Groisman}}\ and\ \bibinfo {author} {\bibfnamefont {V.}~\bibnamefont
  {Steinberg}},\ }\bibfield  {title} {\bibinfo {title} {Elastic turbulence in a
  polymer solution flow},\ }\href@noop {} {\bibfield  {journal} {\bibinfo
  {journal} {Nature}\ }\textbf {\bibinfo {volume} {405}},\ \bibinfo {pages}
  {53} (\bibinfo {year} {2000})}\BibitemShut {NoStop}%
\bibitem [{\citenamefont {Groisman}\ and\ \citenamefont
  {Steinberg}(2001)}]{groisman2001efficient}%
  \BibitemOpen
  \bibfield  {author} {\bibinfo {author} {\bibfnamefont {A.}~\bibnamefont
  {Groisman}}\ and\ \bibinfo {author} {\bibfnamefont {V.}~\bibnamefont
  {Steinberg}},\ }\bibfield  {title} {\bibinfo {title} {Efficient mixing at low
  reynolds numbers using polymer additives},\ }\href@noop {} {\bibfield
  {journal} {\bibinfo  {journal} {Nature}\ }\textbf {\bibinfo {volume} {410}},\
  \bibinfo {pages} {905} (\bibinfo {year} {2001})}\BibitemShut {NoStop}%
\bibitem [{\citenamefont {Traore}\ \emph {et~al.}(2015)\citenamefont {Traore},
  \citenamefont {Castelain},\ and\ \citenamefont
  {Burghelea}}]{traore2015efficient}%
  \BibitemOpen
  \bibfield  {author} {\bibinfo {author} {\bibfnamefont {B.}~\bibnamefont
  {Traore}}, \bibinfo {author} {\bibfnamefont {C.}~\bibnamefont {Castelain}},\
  and\ \bibinfo {author} {\bibfnamefont {T.}~\bibnamefont {Burghelea}},\
  }\bibfield  {title} {\bibinfo {title} {Efficient heat transfer in a regime of
  elastic turbulence},\ }\href@noop {} {\bibfield  {journal} {\bibinfo
  {journal} {Journal of Non-Newtonian Fluid Mechanics}\ }\textbf {\bibinfo
  {volume} {223}},\ \bibinfo {pages} {62–76} (\bibinfo {year}
  {2015})}\BibitemShut {NoStop}%
\bibitem [{\citenamefont {Poole}\ \emph {et~al.}(2012)\citenamefont {Poole},
  \citenamefont {Budhiraja}, \citenamefont {Cain},\ and\ \citenamefont
  {Scott}}]{poole2012emulsification}%
  \BibitemOpen
  \bibfield  {author} {\bibinfo {author} {\bibfnamefont {R.}~\bibnamefont
  {Poole}}, \bibinfo {author} {\bibfnamefont {B.}~\bibnamefont {Budhiraja}},
  \bibinfo {author} {\bibfnamefont {A.}~\bibnamefont {Cain}},\ and\ \bibinfo
  {author} {\bibfnamefont {P.}~\bibnamefont {Scott}},\ }\bibfield  {title}
  {\bibinfo {title} {Emulsification using elastic turbulence},\ }\href@noop {}
  {\bibfield  {journal} {\bibinfo  {journal} {Journal of Non-Newtonian Fluid
  Mechanics}\ }\textbf {\bibinfo {volume} {177}},\ \bibinfo {pages} {15}
  (\bibinfo {year} {2012})}\BibitemShut {NoStop}%
\bibitem [{\citenamefont {Sid}\ \emph {et~al.}(2018)\citenamefont {Sid},
  \citenamefont {Terrapon},\ and\ \citenamefont {Dubief}}]{sid2018two}%
  \BibitemOpen
  \bibfield  {author} {\bibinfo {author} {\bibfnamefont {S.}~\bibnamefont
  {Sid}}, \bibinfo {author} {\bibfnamefont {V.~E.}\ \bibnamefont {Terrapon}},\
  and\ \bibinfo {author} {\bibfnamefont {Y.}~\bibnamefont {Dubief}},\
  }\bibfield  {title} {\bibinfo {title} {Two-dimensional dynamics of
  elasto-inertial turbulence and its role in polymer drag reduction},\ }\href
  {https://doi.org/10.1103/PhysRevFluids.3.011301} {\bibfield  {journal}
  {\bibinfo  {journal} {Phys. Rev. Fluids}\ }\textbf {\bibinfo {volume} {3}},\
  \bibinfo {pages} {011301} (\bibinfo {year} {2018})}\BibitemShut {NoStop}%
\bibitem [{\citenamefont {Robinson}(1991)}]{robinson1991coherent}%
  \BibitemOpen
  \bibfield  {author} {\bibinfo {author} {\bibfnamefont {S.~K.}\ \bibnamefont
  {Robinson}},\ }\bibfield  {title} {\bibinfo {title} {Coherent motions in the
  turbulent boundary layer},\ }\href@noop {} {\bibfield  {journal} {\bibinfo
  {journal} {Annual Review of Fluid Mechanics}\ }\textbf {\bibinfo {volume}
  {23}},\ \bibinfo {pages} {601} (\bibinfo {year} {1991})}\BibitemShut
  {NoStop}%
\bibitem [{\citenamefont {Jim{\'e}nez}\ and\ \citenamefont
  {Pinelli}(1999)}]{jimenez1999autonomous}%
  \BibitemOpen
  \bibfield  {author} {\bibinfo {author} {\bibfnamefont {J.}~\bibnamefont
  {Jim{\'e}nez}}\ and\ \bibinfo {author} {\bibfnamefont {A.}~\bibnamefont
  {Pinelli}},\ }\bibfield  {title} {\bibinfo {title} {{The autonomous cycle of
  near-wall turbulence}},\ }\href@noop {} {\bibfield  {journal} {\bibinfo
  {journal} {J. Fluid Mech.}\ }\textbf {\bibinfo {volume} {389}},\ \bibinfo
  {pages} {335} (\bibinfo {year} {1999})}\BibitemShut {NoStop}%
\bibitem [{\citenamefont {Kolmogorov}(1941)}]{kolmogorov1941local}%
  \BibitemOpen
  \bibfield  {author} {\bibinfo {author} {\bibfnamefont {A.}~\bibnamefont
  {Kolmogorov}},\ }\bibfield  {title} {\bibinfo {title} {The local structure of
  turbulence in incompressible viscous fluid for very large reynolds numbers},\
  }in\ \href@noop {} {\emph {\bibinfo {booktitle} {Dokl. Akad. Nauk SSSR}}},\
  Vol.~\bibinfo {volume} {30}\ (\bibinfo {year} {1941})\ pp.\ \bibinfo {pages}
  {9--13}\BibitemShut {NoStop}%
\bibitem [{\citenamefont {Dubief}\ \emph {et~al.}(2010)\citenamefont {Dubief},
  \citenamefont {White}, \citenamefont {Shaqfeh},\ and\ \citenamefont
  {Terrapon}}]{dubief2010polymer}%
  \BibitemOpen
  \bibfield  {author} {\bibinfo {author} {\bibfnamefont {Y.}~\bibnamefont
  {Dubief}}, \bibinfo {author} {\bibfnamefont {C.~M.}\ \bibnamefont {White}},
  \bibinfo {author} {\bibfnamefont {E.~S.~G.}\ \bibnamefont {Shaqfeh}},\ and\
  \bibinfo {author} {\bibfnamefont {V.~E.}\ \bibnamefont {Terrapon}},\
  }\bibfield  {title} {\bibinfo {title} {{Polymer maximum drag reduction: {A}
  unique transitional state}},\ }in\ \href@noop {} {\emph {\bibinfo {booktitle}
  {Annual Research Briefs}}}\ (\bibinfo {organization} {Center for Turbulence
  Research},\ \bibinfo {address} {Stanford, CA},\ \bibinfo {year} {2010})\ pp.\
  \bibinfo {pages} {395--404}\BibitemShut {NoStop}%
\bibitem [{\citenamefont {Terrapon}\ \emph {et~al.}(2014)\citenamefont
  {Terrapon}, \citenamefont {Dubief},\ and\ \citenamefont
  {Soria}}]{Terrapon2014wu}%
  \BibitemOpen
  \bibfield  {author} {\bibinfo {author} {\bibfnamefont {V.~E.}\ \bibnamefont
  {Terrapon}}, \bibinfo {author} {\bibfnamefont {Y.}~\bibnamefont {Dubief}},\
  and\ \bibinfo {author} {\bibfnamefont {J.}~\bibnamefont {Soria}},\ }\bibfield
   {title} {\bibinfo {title} {{On the role of pressure in elasto-inertial
  turbulence }},\ }\href@noop {} {\bibfield  {journal} {\bibinfo  {journal}
  {Journal of Turbulence}\ }\textbf {\bibinfo {volume} {16}},\ \bibinfo {pages}
  {26} (\bibinfo {year} {2014})}\BibitemShut {NoStop}%
\bibitem [{\citenamefont {Dubief}\ and\ \citenamefont
  {Delcayre}(2000)}]{dubief2000coherent}%
  \BibitemOpen
  \bibfield  {author} {\bibinfo {author} {\bibfnamefont {Y.}~\bibnamefont
  {Dubief}}\ and\ \bibinfo {author} {\bibfnamefont {F.}~\bibnamefont
  {Delcayre}},\ }\bibfield  {title} {\bibinfo {title} {On coherent-vortex
  identification in turbulence},\ }\href@noop {} {\bibfield  {journal}
  {\bibinfo  {journal} {J. of Turbulence}\ }\textbf {\bibinfo {volume} {1}}
  (\bibinfo {year} {2000})}\BibitemShut {NoStop}%
\bibitem [{\citenamefont {Burghelea}\ \emph {et~al.}(2007)\citenamefont
  {Burghelea}, \citenamefont {Segre},\ and\ \citenamefont
  {Steinberg}}]{burghelea2007elastic}%
  \BibitemOpen
  \bibfield  {author} {\bibinfo {author} {\bibfnamefont {T.}~\bibnamefont
  {Burghelea}}, \bibinfo {author} {\bibfnamefont {E.}~\bibnamefont {Segre}},\
  and\ \bibinfo {author} {\bibfnamefont {V.}~\bibnamefont {Steinberg}},\
  }\bibfield  {title} {\bibinfo {title} {Elastic turbulence in von karman
  swirling flow between two disks},\ }\href@noop {} {\bibfield  {journal}
  {\bibinfo  {journal} {Phys. Fluids}\ }\textbf {\bibinfo {volume} {19}},\
  \bibinfo {pages} {053104} (\bibinfo {year} {2007})}\BibitemShut {NoStop}%
\bibitem [{\citenamefont {Shekar}\ \emph {et~al.}(2019)\citenamefont {Shekar},
  \citenamefont {McMullen}, \citenamefont {Wang}, \citenamefont {McKeon},\ and\
  \citenamefont {Graham}}]{ashwin2019critical}%
  \BibitemOpen
  \bibfield  {author} {\bibinfo {author} {\bibfnamefont {A.}~\bibnamefont
  {Shekar}}, \bibinfo {author} {\bibfnamefont {R.~M.}\ \bibnamefont
  {McMullen}}, \bibinfo {author} {\bibfnamefont {S.-N.}\ \bibnamefont {Wang}},
  \bibinfo {author} {\bibfnamefont {B.~J.}\ \bibnamefont {McKeon}},\ and\
  \bibinfo {author} {\bibfnamefont {M.~D.}\ \bibnamefont {Graham}},\ }\bibfield
   {title} {\bibinfo {title} {Critical-layer structures and mechanisms in
  elastoinertial turbulence},\ }\href
  {https://doi.org/10.1103/PhysRevLett.122.124503} {\bibfield  {journal}
  {\bibinfo  {journal} {Phys. Rev. Lett.}\ }\textbf {\bibinfo {volume} {122}},\
  \bibinfo {pages} {124503} (\bibinfo {year} {2019})}\BibitemShut {NoStop}%
\bibitem [{\citenamefont {Dubief}\ \emph {et~al.}(2005)\citenamefont {Dubief},
  \citenamefont {Terrapon}, \citenamefont {White}, \citenamefont {Shaqfeh},
  \citenamefont {Moin},\ and\ \citenamefont {Lele}}]{dubief2005nai}%
  \BibitemOpen
  \bibfield  {author} {\bibinfo {author} {\bibfnamefont {Y.}~\bibnamefont
  {Dubief}}, \bibinfo {author} {\bibfnamefont {V.}~\bibnamefont {Terrapon}},
  \bibinfo {author} {\bibfnamefont {C.}~\bibnamefont {White}}, \bibinfo
  {author} {\bibfnamefont {E.}~\bibnamefont {Shaqfeh}}, \bibinfo {author}
  {\bibfnamefont {P.}~\bibnamefont {Moin}},\ and\ \bibinfo {author}
  {\bibfnamefont {S.}~\bibnamefont {Lele}},\ }\bibfield  {title} {\bibinfo
  {title} {{New answers on the interaction between polymers and vortices in
  turbulent flows}},\ }\href@noop {} {\bibfield  {journal} {\bibinfo  {journal}
  {Flow, turbulence and combustion}\ }\textbf {\bibinfo {volume} {74}},\
  \bibinfo {pages} {311} (\bibinfo {year} {2005})}\BibitemShut {NoStop}%
\bibitem [{\citenamefont {Dubief}\ \emph {et~al.}(2012)\citenamefont {Dubief},
  \citenamefont {Terrapon},\ and\ \citenamefont {Soria}}]{dubief2012analysis}%
  \BibitemOpen
  \bibfield  {author} {\bibinfo {author} {\bibfnamefont {Y.}~\bibnamefont
  {Dubief}}, \bibinfo {author} {\bibfnamefont {V.~E.}\ \bibnamefont
  {Terrapon}},\ and\ \bibinfo {author} {\bibfnamefont {J.}~\bibnamefont
  {Soria}},\ }\bibfield  {title} {\bibinfo {title} {{Analysis of transitional
  polymeric flows and elastic instabilities }},\ }in\ \href@noop {} {\emph
  {\bibinfo {booktitle} {Proceedings of the Summer Program 2012}}}\ (\bibinfo
  {organization} {Center for Turbulence Research},\ \bibinfo {address}
  {Stanford, CA},\ \bibinfo {year} {2012})\ pp.\ \bibinfo {pages}
  {55--63}\BibitemShut {NoStop}%
\bibitem [{\citenamefont {Purnode}\ and\ \citenamefont
  {Legat}(1996)}]{Purnode1996111}%
  \BibitemOpen
  \bibfield  {author} {\bibinfo {author} {\bibfnamefont {B.}~\bibnamefont
  {Purnode}}\ and\ \bibinfo {author} {\bibfnamefont {V.}~\bibnamefont
  {Legat}},\ }\bibfield  {title} {\bibinfo {title} {Hyperbolicity and change of
  type in flows of {FENE-P} fluids},\ }\href@noop {} {\bibfield  {journal}
  {\bibinfo  {journal} {J. Non-Newt. Fluid Mech.}\ }\textbf {\bibinfo {volume}
  {65}},\ \bibinfo {pages} {111 } (\bibinfo {year} {1996})}\BibitemShut
  {NoStop}%
\bibitem [{\citenamefont {Page}\ \emph {et~al.}(2020)\citenamefont {Page},
  \citenamefont {Dubief},\ and\ \citenamefont {Kerswell}}]{page2020traveling}%
  \BibitemOpen
  \bibfield  {author} {\bibinfo {author} {\bibfnamefont {J.}~\bibnamefont
  {Page}}, \bibinfo {author} {\bibfnamefont {Y.}~\bibnamefont {Dubief}},\ and\
  \bibinfo {author} {\bibfnamefont {R.~R.}\ \bibnamefont {Kerswell}},\
  }\bibfield  {title} {\bibinfo {title} {Exact traveling wave solutions in
  viscoelastic channel flow},\ }\href
  {https://doi.org/10.1103/PhysRevLett.125.154501} {\bibfield  {journal}
  {\bibinfo  {journal} {Phys. Rev. Lett.}\ }\textbf {\bibinfo {volume} {125}},\
  \bibinfo {pages} {154501} (\bibinfo {year} {2020})}\BibitemShut {NoStop}%
\bibitem [{\citenamefont {Bird}\ \emph {et~al.}(1987)\citenamefont {Bird},
  \citenamefont {Armstrong},\ and\ \citenamefont
  {Hassager}}]{bird1987dynamics}%
  \BibitemOpen
  \bibfield  {author} {\bibinfo {author} {\bibfnamefont {R.}~\bibnamefont
  {Bird}}, \bibinfo {author} {\bibfnamefont {R.}~\bibnamefont {Armstrong}},\
  and\ \bibinfo {author} {\bibfnamefont {O.}~\bibnamefont {Hassager}},\
  }\href@noop {} {\emph {\bibinfo {title} {{Dynamics of Polymeric Liquids. Vol.
  2: Kinetic Theory}}}}\ (\bibinfo  {publisher} {Wiley-Interscience, 1987,},\
  \bibinfo {year} {1987})\BibitemShut {NoStop}%
\bibitem [{\citenamefont {Varshney}\ and\ \citenamefont
  {Steinberg}(2018)}]{varshney2018drag}%
  \BibitemOpen
  \bibfield  {author} {\bibinfo {author} {\bibfnamefont {A.}~\bibnamefont
  {Varshney}}\ and\ \bibinfo {author} {\bibfnamefont {V.}~\bibnamefont
  {Steinberg}},\ }\bibfield  {title} {\bibinfo {title} {Drag enhancement and
  drag reduction in viscoelastic flow},\ }\href@noop {} {\bibfield  {journal}
  {\bibinfo  {journal} {Physical Review Fluids}\ }\textbf {\bibinfo {volume}
  {3}},\ \bibinfo {pages} {103302} (\bibinfo {year} {2018})}\BibitemShut
  {NoStop}%
\bibitem [{\citenamefont {Choueiri}\ \emph {et~al.}(2018)\citenamefont
  {Choueiri}, \citenamefont {Lopez},\ and\ \citenamefont
  {Hof}}]{choueiri2018exceeding}%
  \BibitemOpen
  \bibfield  {author} {\bibinfo {author} {\bibfnamefont {G.~H.}\ \bibnamefont
  {Choueiri}}, \bibinfo {author} {\bibfnamefont {J.~M.}\ \bibnamefont
  {Lopez}},\ and\ \bibinfo {author} {\bibfnamefont {B.}~\bibnamefont {Hof}},\
  }\bibfield  {title} {\bibinfo {title} {Exceeding the asymptotic limit of
  polymer drag reduction},\ }\href@noop {} {\bibfield  {journal} {\bibinfo
  {journal} {Physical review letters}\ }\textbf {\bibinfo {volume} {120}},\
  \bibinfo {pages} {124501} (\bibinfo {year} {2018})}\BibitemShut {NoStop}%
\bibitem [{\citenamefont {Tamano}\ \emph {et~al.}(2009)\citenamefont {Tamano},
  \citenamefont {Itoh}, \citenamefont {Hotta}, \citenamefont {Yokota},\ and\
  \citenamefont {Morinishi}}]{tamano2009effect}%
  \BibitemOpen
  \bibfield  {author} {\bibinfo {author} {\bibfnamefont {S.}~\bibnamefont
  {Tamano}}, \bibinfo {author} {\bibfnamefont {M.}~\bibnamefont {Itoh}},
  \bibinfo {author} {\bibfnamefont {S.}~\bibnamefont {Hotta}}, \bibinfo
  {author} {\bibfnamefont {K.}~\bibnamefont {Yokota}},\ and\ \bibinfo {author}
  {\bibfnamefont {Y.}~\bibnamefont {Morinishi}},\ }\bibfield  {title} {\bibinfo
  {title} {Effect of rheological properties on drag reduction in turbulent
  boundary layer flow},\ }\href@noop {} {\bibfield  {journal} {\bibinfo
  {journal} {Phys. Fluids}\ }\textbf {\bibinfo {volume} {21}},\ \bibinfo
  {pages} {055101} (\bibinfo {year} {2009})}\BibitemShut {NoStop}%
\bibitem [{\citenamefont {Garg}\ \emph {et~al.}(2018)\citenamefont {Garg},
  \citenamefont {Chaudhary}, \citenamefont {Khalid}, \citenamefont {Shankar},\
  and\ \citenamefont {Subramanian}}]{garg2018viscoelastic}%
  \BibitemOpen
  \bibfield  {author} {\bibinfo {author} {\bibfnamefont {P.}~\bibnamefont
  {Garg}}, \bibinfo {author} {\bibfnamefont {I.}~\bibnamefont {Chaudhary}},
  \bibinfo {author} {\bibfnamefont {M.}~\bibnamefont {Khalid}}, \bibinfo
  {author} {\bibfnamefont {V.}~\bibnamefont {Shankar}},\ and\ \bibinfo {author}
  {\bibfnamefont {G.}~\bibnamefont {Subramanian}},\ }\bibfield  {title}
  {\bibinfo {title} {Viscoelastic pipe flow is linearly unstable},\ }\href@noop
  {} {\bibfield  {journal} {\bibinfo  {journal} {Physical review letters}\
  }\textbf {\bibinfo {volume} {121}},\ \bibinfo {pages} {024502} (\bibinfo
  {year} {2018})}\BibitemShut {NoStop}%
\bibitem [{\citenamefont {Chaudhary}\ \emph {et~al.}(2020)\citenamefont
  {Chaudhary}, \citenamefont {Garg}, \citenamefont {Subramanian},\ and\
  \citenamefont {Shankar}}]{chaudhary2020linear}%
  \BibitemOpen
  \bibfield  {author} {\bibinfo {author} {\bibfnamefont {I.}~\bibnamefont
  {Chaudhary}}, \bibinfo {author} {\bibfnamefont {P.}~\bibnamefont {Garg}},
  \bibinfo {author} {\bibfnamefont {G.}~\bibnamefont {Subramanian}},\ and\
  \bibinfo {author} {\bibfnamefont {V.}~\bibnamefont {Shankar}},\ }\bibfield
  {title} {\bibinfo {title} {Linear instability of viscoelastic pipe flow},\
  }\href@noop {} {\bibfield  {journal} {\bibinfo  {journal} {arXiv preprint
  arXiv:2003.09369}\ } (\bibinfo {year} {2020})}\BibitemShut {NoStop}%
\bibitem [{\citenamefont {Berti}\ \emph {et~al.}(2008)\citenamefont {Berti},
  \citenamefont {Bistagnino}, \citenamefont {Boffetta}, \citenamefont
  {Celani},\ and\ \citenamefont {Musacchio}}]{berti2008two}%
  \BibitemOpen
  \bibfield  {author} {\bibinfo {author} {\bibfnamefont {S.}~\bibnamefont
  {Berti}}, \bibinfo {author} {\bibfnamefont {A.}~\bibnamefont {Bistagnino}},
  \bibinfo {author} {\bibfnamefont {G.}~\bibnamefont {Boffetta}}, \bibinfo
  {author} {\bibfnamefont {A.}~\bibnamefont {Celani}},\ and\ \bibinfo {author}
  {\bibfnamefont {S.}~\bibnamefont {Musacchio}},\ }\bibfield  {title} {\bibinfo
  {title} {Two-dimensional elastic turbulence},\ }\href@noop {} {\bibfield
  {journal} {\bibinfo  {journal} {Physical Review E}\ }\textbf {\bibinfo
  {volume} {77}},\ \bibinfo {pages} {055306} (\bibinfo {year}
  {2008})}\BibitemShut {NoStop}%
\bibitem [{\citenamefont {Berti}\ and\ \citenamefont
  {Boffetta}(2010)}]{berti2010elastic}%
  \BibitemOpen
  \bibfield  {author} {\bibinfo {author} {\bibfnamefont {S.}~\bibnamefont
  {Berti}}\ and\ \bibinfo {author} {\bibfnamefont {G.}~\bibnamefont
  {Boffetta}},\ }\bibfield  {title} {\bibinfo {title} {Elastic waves and
  transition to elastic turbulence in a two-dimensional viscoelastic kolmogorov
  flow},\ }\href@noop {} {\bibfield  {journal} {\bibinfo  {journal} {Physical
  Review E}\ }\textbf {\bibinfo {volume} {82}},\ \bibinfo {pages} {036314}
  (\bibinfo {year} {2010})}\BibitemShut {NoStop}%
\bibitem [{\citenamefont {Jun}\ and\ \citenamefont
  {Steinberg}(2009)}]{jun2009power}%
  \BibitemOpen
  \bibfield  {author} {\bibinfo {author} {\bibfnamefont {Y.}~\bibnamefont
  {Jun}}\ and\ \bibinfo {author} {\bibfnamefont {V.}~\bibnamefont
  {Steinberg}},\ }\bibfield  {title} {\bibinfo {title} {Power and pressure
  fluctuations in elastic turbulence over a wide range of polymer
  concentrations},\ }\href@noop {} {\bibfield  {journal} {\bibinfo  {journal}
  {Phys. Rev. Lett.}\ }\textbf {\bibinfo {volume} {102}},\ \bibinfo {pages}
  {124503} (\bibinfo {year} {2009})}\BibitemShut {NoStop}%
\bibitem [{\citenamefont {Choueiri}\ \emph {et~al.}(2021)\citenamefont
  {Choueiri}, \citenamefont {Lopez}, \citenamefont {Varshney}, \citenamefont
  {Sankar},\ and\ \citenamefont {Hof}}]{choueiri2021experimental}%
  \BibitemOpen
  \bibfield  {author} {\bibinfo {author} {\bibfnamefont {G.~H.}\ \bibnamefont
  {Choueiri}}, \bibinfo {author} {\bibfnamefont {J.~M.}\ \bibnamefont {Lopez}},
  \bibinfo {author} {\bibfnamefont {A.}~\bibnamefont {Varshney}}, \bibinfo
  {author} {\bibfnamefont {S.}~\bibnamefont {Sankar}},\ and\ \bibinfo {author}
  {\bibfnamefont {B.}~\bibnamefont {Hof}},\ }\bibfield  {title} {\bibinfo
  {title} {Experimental observation of the origin and structure of
  elasto-inertial turbulence},\ }\href@noop {} {\bibfield  {journal} {\bibinfo
  {journal} {arXiv preprint arXiv:2103.00023}\ } (\bibinfo {year}
  {2021})}\BibitemShut {NoStop}%
\end{thebibliography}%


%

\section{Appendix}
\renewcommand{\thefigure}{A\arabic{figure}}

\setcounter{figure}{0}

\setcounter{table}{0}
\renewcommand{\thetable}{A\arabic{table}}

\subsection{Grid resolution study}
\begin{table}[h]
    \centering
    \begin{tabular}{cccccc}
    $Nx$ & $Nz$ & $\Delta_{z,min}/h$ & $\overline{f}(t)$ & $\mathrm{RMS}(f)$ & Regime \\
    \hline
      128     & 129 & $10^{-4}$ & $7.1\times10^{-3}$ & $9.4\times10^{-4}$ & IAR   \\
      256    & 257 & $10^{-4}$ & $6.8\times10^{-3}$ & $9.3\times10^{-4}$ & IAR \\
      512    & 513 & $10^{-3}$ & $8.2\times10^{-3}$ & $4.1\times10^{-3}$ & CAR\\
      512    & 513 & $5\times10^{-4}$ & $7.6\times10^{-3}$ & $3.1\times10^{-3}$ & CAR\\
      \hline
      512    & 513 & $10^{-4}$ & $7.6\times10^{-3}$ & $3.16\times10^{-3}$ & CAR \\
      \hline
      512    & 513 & $5\times10^{-5}$ & $7.6\times10^{-3}$ & $3.15\times10^{-3}$ & CAR\\
      1024    & 1025 & $10^{-4}$ & $7.6\times10^{-3}$ & $3.15\times10^{-3}$ & CAR\\
      \hline
    \end{tabular}
    \caption{Example of numerical resolution study for regime CAR defined in Fig.~\ref{fig:1}. All simulations in this case were performed at $Re_b=1000$ in a domain $L_x\times L_z=2\pi h\times 2h$. The monitoring metrics are the mean pressure gradient $f$ and its RMS. The resolution used for production is framed by two horizontal lines.} 
    \label{tab:numres}
\end{table}
Appropriate grid resolution is critical to the simulation of the different states of EIT. In the course of the present and previous studies, we identified three parameters of crucial importance: Overall resolution $N_x\times Nz$, size of the cell at the wall $\Delta z_{min}$ and Schmidt number. In this section, the study of these parameters is illustrated  by a fully chaotic flow (CR) obtained with $Re_b=1000$, $L=50$, $Wi=50$ and $\beta = 0.9$. The characteritics of the simulations used in this grid convergence studies are compiled in Table~\ref{tab:numres}, which also reports the mean and RMS of the pressure gradient $f(t)$ driving the constant mass flow (see Eq.~\ref{eq:mom}, and the flow regime.

\subsection{Effects of grid resolution, and Schmidt number of small scales}
\begin{figure}[h]

\centerline{\includegraphics[width=
\columnwidth]{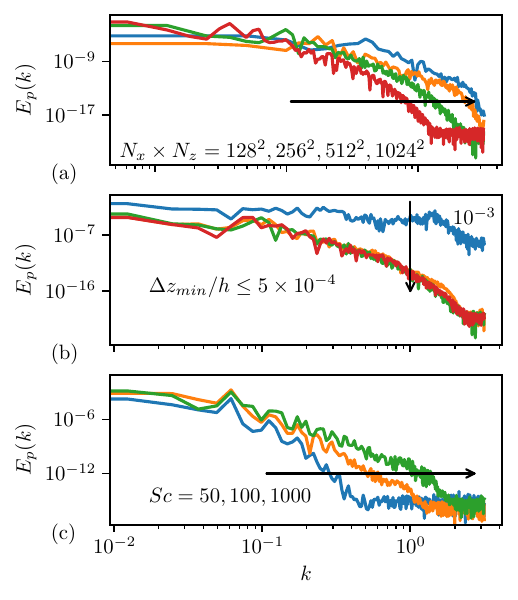}}

\caption{\label{fig:A0} Spectral analysis of the resolution study for $Re_b=1000$, $L=50$, $Wi=100$, $\beta=0.9$. (a) shows the effect of resolution on the fluctuations of wall pressure, (b), the effect of the size of mesh at the wall, and (c) the effect of the Schmidt number.}
\end{figure}
As discussed in Dubief \textit{et al.}\cite{dubief2005nai}, there is no dissipative mechanism in the exact FENE-P equation. The high-order compact scheme used for the advection terms introduces numerical dissipation at high wavenumbers \cite{dubief2005nai} necessary to avoid a buildup of energy at small scales due to the hyperbolic nature of FENE-P. Consequently gradients of $C_{ij}$ and $T_{ij}$  are expected to become sharper with higher resolutions. Through Eq.~\ref{eq:ddp}, larger gradients impact the pressure distribution throughout the domain. Fig~\ref{fig:A0}(a) shows the power spectral densities in the streamwise direction of wall pressure fluctuations for increasing resolution with fixed $\Delta z_{min}/h=10^{-4}$ and $Sc=1000$. The two lowest resolutions, $128^2$ and $256^2$ underestimate the spectral content of pressure fluctuations across scales, and as reported in Table~\ref{tab:numres}, they yield a flow  regime different than the regime simulated for the two highest resolutions. 

  Another critical parameter is the smallest grid size $\Delta_{z,min}$ at the wall. A parametric study established that $\Delta_{z,min}=10^{-4}$ is necessary to capture the intense gradient of polymer stress at the wall in chaotic regimes. Fig.~\ref{fig:A0}b highlights the spurious increase of energy in the PSD of wall pressure fluctuations across all scales caused by too coarse of a resolution ($\Delta_{z,min}=10^{-3}$), and the convergence of the spectra for ($\Delta_{z,min}=5\times10^{-4},10^{-4},5\times10^{-5}$)

The Schmidt number study (Fig.~\ref{fig:A0}c) shows an expected reduction of the energy content at small scale with decreasing Schmidt number. Note that  decreasing the Schmidt number is not found to change the flow regime unless $Sc\lesssim 5$, in which case the flow becomes laminar.

The simulation of EIT is therefore a necessary compromise between aiming for the highest resolution for accuracy at the maximum number of wavenumbers and  the ability to run long simulations to capture low frequency behavior in statistics (like IAR). It is also important to note that the stiffness of the FENE-P equations requires numerical artifacts, such as GAD, upwind schemes. The numerical methods bring certain levels of numerical dispersion and dissipation into the solution. The following section demonstrates that SAR is not influenced by numerical dispersion and dissipation. The effects of numerical dissipation on EIT is addressed in the previous and current sections in terms of the resolution study. The higher the resolution, the lower the contribution of numerical dissipation is. The influence of dispersion is also small as shown by the resolution study. Nonetheless any research focused on finding the exact bounds of the CR, CAR, IAR regimes should consider quantifying the uncertainty from numerical dispersion. This is beyond the scope of this paper.

\subsection{Effects of numerical dissipation and dispersion on SAR}
\begin{figure}[]

\centerline{\includegraphics[width=
\columnwidth]{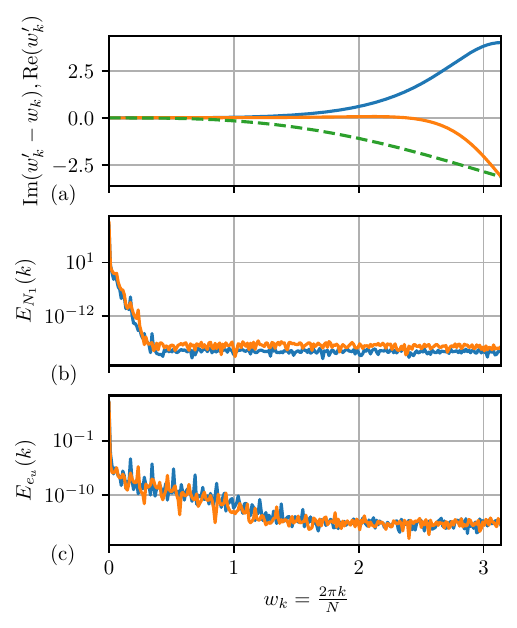}}

\caption{\label{fig:A1} (a) Imaginary (negative curves) and real (positive curve) components of the modified wavenumbers for the upwind compact scheme \cite{dubief2005nai} (solid lines) and the staggered second order scheme (dashed) for the spatial first derivative. The imaginary and real components quantify the numerical dispersion and dissipation, respectively, of each scheme as a function of the wavenumber. (b) Power spectra of first normal stress difference at the wall of two SAR simulations ($L=100,200$, $Wi=100$, $\beta=0.9$). (c)  Power spectra of turbulent kinetic energy for the same simulations. }
\end{figure}
Lastly, the power spectra of $N_1$ are shown in Fig.~\ref{fig:A1}a for two SAR regimes. Their distribution as a function of the wavenumber is compared to the numerical dissipation and dispersion of the upwind compact scheme used for the advection term \cite{dubief2005nai} in the FENE-P equation (Eq.~\ref{eq:C}). In the Fourier space, the exact first derivative of a function $f$ defined on $k\in[1,N]$ computational nodes is 
\begin{equation*}
    \left.\widehat{\pder{x}{f}}\right\vert_k = \hat{\imath}w_k\hat{f}_k\,,
\end{equation*} 
where $w_k=2\pi k/N$ is the wavenumber vector, $\hat{\imath}^2=-1$ and $\hat{f}_k$ is the Fourier coefficient vector of $f$. Any finite difference or finite volume scheme for first derivative can be recast in the Fourier space as 
\begin{equation*}
    \left.\widehat{\pder{x}{f}}\right\vert_k = \hat{\imath}w'_k\hat{f}_k\,,
\end{equation*} 
where $w'_k$ is the modified wavenumber of the numerical scheme, here the upwind compact scheme defined in \cite{dubief2005nai}. The real component $Re(w'_k)$ of the modified wavenumber represents the numerical dissipation at wavenumber $w_k$ of the scheme. The motivation for an upwind compact scheme is to confine numerical dissipation in the high wavenumbers in order to damp Gibbs oscillations near the grid cutoff. Gibbs oscillations are caused by the presence of large gradients of $C_{ij}$ resulting from the stiffness of the FENE-P model.

Unlike spectral methods, any finite difference (FD) scheme for derivative introduces a certain level of numerical dispersion. The numerical dispersion as a function of the wavenumber is defined as the $\mathrm{Re}(w'_k)$. Central FD schemes are non-dissipative, however the upwind compact scheme is, by design. The numerical dissipation is the real part of the modified wavenumber, $\mathrm{Im}(w'_k-w_k)$. Fig.~\ref{fig:A1}a shows the dispersion of staggered second-order FD scheme used for the advection of momentum and velocity divergence, and the upwind compact scheme used for the advection of $C_{ij}$, as well as the numerical dissipation of the upwind compact scheme. The departure from zero for the numerical dispersion and dissipation of each scheme is compared to the spectra of the first normal stress difference ($N_1$) and spectra of turbulent kinetic energy for two simulations of SAR. The relevant dynamic scales for $N_1$ are confined to wavenumber smaller than the range of wavenumbers experiencing numerical dissipation ($w_k\gtrsim 1.5$) and much smaller that the range of wavenumbers experiencing dispersion ($w_k\gtrsim 2.5$) for the compact upwind scheme. The numerical dispersion of the staggered central FD scheme affects a range of scales whose energy 10 decades or more smaller than the large scale energy of turbulent kinetic energy. Note that the dispersion of the central compact scheme used in the momentum transport equation (Eq.~\ref{eq:mom}) is not shown in the graph but comparable to that of the upwind compact scheme. 

It should be noted that SAR simulations were successfully run with $Sc=50$, which is not surprising considering the absence of small scales shown in Figs~\ref{fig:A1}b-c.

\end{document}